\documentclass[11pt,tightenlines,eqsecnum,aps,amsmath,amssymb,nofootinbib,prd]{revtex4}
\usepackage[english]{babel}
\usepackage[utf8]{inputenc}
\usepackage{amsmath}
\usepackage{amssymb}
\usepackage{graphicx}
\usepackage{subcaption}
\usepackage{hyperref}
\usepackage{float}
\usepackage{tikz}
\usepackage[margin=1cm]{caption}
\usetikzlibrary{quotes,angles}
\usepackage{titlesec}
\usepackage{lscape}
\usetikzlibrary{shapes.geometric, arrows}
\usetikzlibrary{positioning}

\newcommand\StepSubequations{
  \stepcounter{parentequation}
  \gdef\theparentequation{\arabic{parentequation}}
  \setcounter{equation}{0}
}

\tikzstyle{startstop} = [rectangle, rounded corners, minimum width=3cm, minimum height=1cm,text centered, draw=black, fill=blue!30]
\tikzstyle{init} = [rectangle, minimum width=3cm, minimum height=1cm, text width=3cm, text centered, draw=black, fill=red!30]
\tikzstyle{evo} = [rectangle, minimum width=3cm, minimum height=1cm, text width=3cm, text centered, draw=black, fill=yellow!30]
\tikzstyle{amp} = [rectangle, minimum width=3cm, minimum height=1cm, text width=3cm, text centered, draw=black, fill=green!30]
\tikzstyle{deci} = [diamond, text centered, draw=black, fill=blue!30]
\tikzstyle{arrow} = [thick,->,>=stealth]

\begin{document}
	
	\title{Efficient fully precessing gravitational waveforms for binaries with neutron stars}
	
	\author{Michael LaHaye}
	\affiliation{Department of Physics, University of Guelph, Guelph, Ontario, N1G 2W1, Canada}
	\author{Huan Yang}
	\affiliation{Perimeter Institute for Theoretical Physics, Ontario, N2L 2Y5, Canada}
	\affiliation{Department of Physics, University of Guelph, Guelph, Ontario, N1G 2W1, Canada}
	\author{B\'{e}atrice Bonga}
	\affiliation{Institute for Mathematics, Astrophysics and Particle Physics,Radboud University, 6525 AJ Nijmegen, The Netherlands}
	\author{Zhenwei Lyu}
	\affiliation{Kavli Institute for Astronomy and Astrophysics, Peking University, Beijing 100871, China}
	\affiliation{Department of Physics, University of Guelph, Guelph, Ontario, N1G 2W1, Canada}
	
	\begin{abstract}
		We construct an efficient frequency domain waveform for generic circular compact object binaries that include neutron stars. The orbital precession is solved on the radiation reaction timescale (and then transformed to the frequency domain), which is used to map the non-precessional waveform from the source frame of the binary to the lab frame. The treatment of orbital precession is different from that for precessional binary black holes, as $\chi_{\rm eff}$ is no longer conserved due to the spin-induced quadrupole moments of neutron stars. We show that the new waveform achieves $\le 10^{-4}$ mismatch compared with waveforms generated by numerically evolved precession for neutron star-black hole systems for $\ge 90\%$ configurations with component mass/spin magnitude assumed in the analysis and randomized initial spin directions. We expect this waveform to be useful to test the nature of the mass-gap objects similar to the one discovered in GW 190814 by measuring their spin-induced quadrupole moments, as it is possible that these mass-gap objects are rapidly spinning. It is also applicable for the tests of black hole mimickers in precessional binary black hole events, if the black hole mimicker candidates have nontrivial spin-induced quadrupole moments.
	\end{abstract}

	\maketitle
	
	\section{Introduction}
	Since the first gravitational wave (GW) observation in 2015,  almost a hundred binary coalescences have been detected \cite{LIGOScientific:2021djp,LIGOScientific:2020ibl,LIGOScientific:2018mvr}, including  black hole binaries (BHBH), neutron star binaries (NSNS) and neutron star-black hole binaries (NSBH).
	In order to distinguish neutron stars from black holes in these binaries, besides information from electromagnetic counterparts, a natural way is to measure the tidal Love number of the compact object \cite{Flanagan:2007ix}, which is zero for black holes but could be $\mathcal{O}(10^2-10^3)$ for normal neutron stars.  However, no definite detection of nonzero Love number has been made yet. For example, in the first binary neutron star event detected (GW 170817), the dimensionless tidal Love number is constrained to be $\le 800$ \cite{LIGOScientific:2017vwq,Radice:2017lry}. In addition, the tidal Love number decreases sharply with increasing neutron star mass. For heavy neutron stars near their maximum mass, the corresponding dimensionless Love number is likely only $\mathcal{O}(1)$ \cite{Hinderer:2007mb, Hu:2020ujf} , which is at best detectable by the third-generation gravitational wave detectors \cite{Castro:2022mpw}.
	Without definite information from tidal Love numbers and electromagnetic counterparts, a common practice is to label  compact objects with mass $\le 3 M_\odot$ as neutron stars, and those with mass $\ge 5 M_\odot$ as black holes. This is motivated by the lack of black hole observation from X-ray binaries (the ``mass-gap") and neutron star equation-of-state considerations. However, it remains an open question whether there is a detectable population of low-mass black holes (that is, black holes with comparable masses to neutron stars) \cite{Yang:2017gfb}, possibly coming from delayed supernovae explosions, binary neutron star mergers or primordial black holes. A recent study suggests that they may also take place in extreme mass ratio inspirals suitable for space-borne gravitational wave detections \cite{Pan:2021lyw}, with an accelerated formation rate through the interaction with accretion disks \cite{Pan:2021oob,Pan:2021ksp}. It is both physically interesting and astrophysically important to unambiguously identify the nature of (at least some of the) mass-gap objects. \\
		

	The measurement of the spin-induced quadrupole moment may provide a promising method to distinguish neutron stars from low-mass black holes in cases in which this may not be  possible otherwise.  A key example being GW190814 \cite{LIGOScientific:2020zkf}, which describes the coalescence of a $\sim 23 M_\odot$ black hole with a compact object of $\sim 2.6 M_\odot$; this compact object may be a low-mass black hole or a heavy neutron star \cite{Vattis:2020iuz,Godzieba:2020tjn,Most:2020bba} \footnote{If more exotic objects such as boson stars exist, they may be candidates of mass-gap objects as well with the corresponding tidal Love number potentially measurable \cite{Mendes:2016vdr,Cardoso:2019rvt}.}. With the lighter object lying squarely inside the lower mass gap, an extremely small tidal Love number (expected to be $\le \mathcal{O}(1)$) and no optical counterpart, it is difficult to determine the identity of this object. A measurement of the spin-induced quadrupole moment from the gravitational waveform  could provide insight into the nature of this object, if sufficiently different from that of a black hole.  This particular event is especially tantalizing: While neutron stars are not in general expected to have large spins, if this object were a neutron star, it is natural to expect that it would be rapidly spinning in order to support its large mass.  On the other hand, if it were a low-mass black hole, large spin is also expected if it is formed in a binary neutron star merger or delayed supernova with significant accretion. This potentially large spin would make the effects of spin-precession more pronounced, making it a good candidate for the measurement of the spin-induced quadrupole moment. In particular, we can write the spin-induced quadrupole moment as
	\begin{align}
	Q = \kappa a^2 m^3
	\end{align}
	with $Q$ being the magnitude of the quadrupole moment, $m$ its mass and $a$ its spin. For black holes, the quadrupole constant $\kappa$ equals one (i.e., the Kerr metric) and for  neutron stars $\kappa$ is a (larger-than-one) number depending on the star's mass and equation of state (see Fig.~1 of \cite{Harry:2018hke}).\\
	
	In order to determine the spin-induced quadrupole moment, accurate waveform models are required. However, currently, there are no efficient methods of generating waveforms that include the effects of the spin-induced quadrupole moment on the precession of generic compact objects. Such methods only exist for black hole binaries \cite{Chatziioannou:2017tdw, Klein:2021jtd,Chatziioannou:2016ezg}, but these have not been extended to include the effects of the quadrupole moment of neutron stars.  This is the goal of this paper.  We solve the spin dynamics component of the waveform, other parts of the waveform generation are then kept the same as non precessional systems.\\

	For precessing compact binaries, it is computationally expensive to track the binary evolution on the precession timescale, which is generally longer than the orbital timescale but shorter than the radiation reaction timescale. For binary black hole systems, it has been shown that there are a sufficient number of conserved quantities (in particular, $\chi_{\rm eff}$, as discussed later) such that the spin evolution equation can be solved algebraically after performing an average over the precession timescale \cite{Chatziioannou:2017tdw, Klein:2021jtd}. However, the quantity $\chi_{\rm eff}$ is no longer conserved for generic black hole-neutron star systems as neutron stars have $\kappa \neq 1$, so that the precessional binary black hole waveform no longer applies for binaries with neutron stars, even without considering tidal effects. \\
	
	
	The separation of the orbital, precession, and radiation reaction timescales has been used in much of the literature starting with \cite{Apostolatos:1994mx} and more recently \cite{Kesden:2014sla,Gerosa:2015tea}. We will work within the orbit averaged equations, and take advantage of the separation between the precession and radiation reaction (RR) timescales to separate our description into a secular portion (determined by the RR) and a periodic portion (determined by the conservative dynamics).  This process requires first solving the conservative problem, so we start by finding an approximate analytic solution to the conservative dynamics.  When solving the conservative problem we have three spins each with three components, so that we have nine variables describing the system.  For black hole binaries there are seven conserved quantities and two dynamical variables.  As mentioned, one of these conserved quantities, $\chi_{\rm eff}$, is no longer conserved when one of the compact objects is not a black hole.  This leaves us with a choice when solving the conservative dynamics: either find a new conserved quantity to reduce the problem back to two dynamical variables or work with three dynamical variables.\\
	
	To decide which is the preferable option we looked at the BHBH case for insight. For black hole binaries, the conservative dynamics have been solved for several different choices of dynamical variables \cite{Chatziioannou:2017tdw, Klein:2021jtd}. We follow the definitions used in Klein (2021) \cite{Klein:2021jtd}, namely using the sum ($\chi_{\text{eff}}$) and difference ($\delta\chi$) of the spins projected onto the orbital angular momentum as two of our three dynamical variables.  For this choice,  the conservative problem has an exact solution, but this solution relies on the specific form of the evolution equation for $\delta\chi$: because it is cubic, its roots can be found exactly making the solution efficient. We find that the new conserved quantity in the general case is quadratic in both $\chi_{\text{eff}}$ and $\delta\chi$. This changes the cubic form of $\delta\chi$'s equation so that we can no longer use the techniques that applied previously to solve the system.  Thus, while using the conserved quantity directly to reduce the number of dynamical variables may be tempting at first, we opt to continue working with three dynamical variables.\\
	
	With this choice in mind we find an approximate analytic solution to the conservative dynamics, which can be described in terms of an average, amplitude, and precession phase.  The average (secular) portion and amplitude both evolve on the RR timescale. The precession phase obviously evolves on the precession timescale, however, it can be evolved accurately on the RR timescale because the frequency only evolves on the RR timescale.  As a result, one can evolve all quantities on the RR timescale, while still fully accounting for the precession.  \\
	
	The organization of the paper is as follows.  In Sec.~\ref{app:variables} we lay out the relevant definitions, and in the following section, Sec.~\ref{sec:eom}, we lay out the relevant equations.  In Sec.~\ref{sec:analytic-sol}, we solve the conservative dynamics, and in Sec.~\ref{sec:radiation-reaction} we introduce radiation reaction.  In Sec.~\ref{sec:comparison} we evaluate the accuracy of this model, and compare the waveform with the binary black hole case to illustrate the phase difference introduced by the spin-induced quadrupole moment.
	
	
	\section{Angular momenta and spin variables}
	\label{app:variables}
	
	A note about notation: we will use an arrow $\vec{V}$  to denote vectors, and a hat $\hat{V}$ to denote unit vectors, and merely the symbol itself $V$ to denote the magnitudes.  From the total mass $M = m_1 + m_2$, we can define the reduced masses	
	\begin{equation}\label{eq:eq:S00B-SS001-EQ001}
		\begin{split}
			\mu_i &= \frac{m_i}{M}.
		\end{split}
	\end{equation}	
	From this, it will be useful to define two combinations of the two reduced masses, the first being their difference
	\begin{equation}\label{eq:eq:S00B-SS001-EQ002}
		\begin{split}
			\delta\mu = \mu_1 - \mu_2,
		\end{split}
	\end{equation}	
	which is small in the equal mass limit.  The second combination is their product (the symmetric mass ratio)	
	\begin{equation}\label{eq:eq:S00B-SS001-EQ003}
		\begin{split}
			\eta &= \mu_1 \mu_2.
		\end{split}
	\end{equation}	
	The magnitude of the orbital angular momentum can be related to the PN parameter through this symmetric mass ratio	
	\begin{equation}\label{eq:eq:S00B-SS001-EQ004}
		\begin{split}
			L &= \frac{\eta}{y}.
		\end{split}
	\end{equation}	
	This is one of the conserved quantities of the conservative dynamics. The next quantities of interest are the dimensionless spin parameter	
	\begin{equation}\label{eq:eq:S00B-SS001-EQ005}
		\begin{split}
			\chi_i = \frac{S_i}{{m_i}^2}.
		\end{split}
	\end{equation}	
	and the reduced spins	
	\begin{equation}\label{eq:eq:S00B-SS001-EQ006}
		\begin{split}
			\vec{s}_i = \frac{1}{\mu_i} \vec{S}_i \; .
		\end{split}
	\end{equation}
	
	The magnitudes of the spins, $S_1$ and $S_2$ are two more conserved quantities on the precession timescale.  From the reduced spins and the orbital angular momentum, the total angular momentum is defined as
		\begin{equation}\label{eq:eq:S00B-SS001-EQ007}
		\begin{split}
			\vec{J} = \vec{L} + \mu_1 \vec{s}_1 + \mu_2 \vec{s}_2 \; .
		\end{split}
	\end{equation}
	Its three components form three additional conserved quantities of the conservative dynamics (and consequently its magnitude is another, but it is not unique).\\
	
	These spins evolve according to the precession equations
	\begin{equation}\label{eq:eq:S00B-SS001-EQ008}
		\begin{split}
			\frac{d \hat{L} }{dt} &= -y^6 (\Omega_1 + \Omega_2)\\
			\frac{d \vec{s}_1 }{dt} &= \mu_2 y^5 \Omega_1\\
			\frac{d \vec{s}_2 }{dt} &= \mu_1 y^5 \Omega_2
		\end{split}
	\end{equation}
	where 
	\begin{equation}\label{eq:eq:S00B-SS001-EQ009}
		\begin{split}
			\Omega_i = \left\{ \frac{1}{2}\mu_i + \frac{3}{2} \left[ 1 - y \hat{L} \cdot (\kappa_i \vec{s}_i + \vec{s}_j ) \right] \right\} \hat{L} \times \vec{s}_i + \frac{1}{2} y \vec{s}_j \times \vec{s}_i \; .
		\end{split}
	\end{equation}	
	
	From these quantities, we denote the sum of the projections of the reduced spins onto the orbital angular momentum as $\chi_{\rm eff}$
	\begin{equation}\label{eq:eq:S00B-SS001-EQ011}
		\begin{split}
			\chi_{\text{eff}} = \hat{L} \cdot \vec{s}_1 + \hat{L} \cdot \vec{s}_2,
		\end{split}
	\end{equation}
	
	which is one of the conserved quantities of the BBH system first found by \cite{Damour:2001tu} and later shown to be conserved by \cite{Racine:2008qv} (after orbit averaging, it is not instantaneously conserved).  The difference is $\delta \chi$
	\begin{equation}\label{eq:eq:S00B-SS001-EQ012}
		\begin{split}
			\delta\chi = \hat{L} \cdot \vec{s}_1 - \hat{L} \cdot \vec{s}_2.
		\end{split}
	\end{equation}
	
	The angle between $\hat{L}$ and $\hat{J}$ is denoted	
	\begin{equation}\label{eq:eq:S00B-SS001-EQ013}
		\begin{split}
			\cos(\theta_L) = \hat{L} \cdot \hat{J},
		\end{split}
	\end{equation}	
	which can also be written in terms of the other projections using Eq.~\eqref{eq:eq:S00B-SS001-EQ007}-\eqref{eq:eq:S00B-SS001-EQ012}	
	\begin{equation}\label{eq:eq:S00B-SS001-EQ014}
		\begin{split}
			\cos(\theta_L) = \frac{1}{2J} (2L + \chi_{\text{eff}} + \delta\mu \delta\chi) \; .
		\end{split}
	\end{equation}
	
	\section{Equations of motion}\label{sec:eom}
	The precession equations for compact objects on circular orbits including leading post-Newtonian (PN) order spin-orbit and spin-spin interactions --- without radiation reaction terms --- in principle require nine variables for the description: three for the spin of each body and three for the Newtonian angular momentum.  At 2.5 post-Newtonian order one can show that there are six straightforward conserved quantities (the magnitudes of each individual angular momentum $S_1$, $S_2$, and $L$ as well as the three components of the total angular momentum).  As a result, we will need three additional variables to complete the description of the angular momenta.\\
	
	When the two compact objects are black holes, there is in fact an easily identifiable seventh conserved quantity, $\chi_{\rm eff}$, which is a projection of the sum of the reduced spins onto the direction of the Newtonian orbital angular momentum $L$ (see Appendix. \ref{app:variables} for the mathematical expression).  This would leave two dynamical quantities to describe the system completely.  Previously for the binary black hole scenario, the square of the magnitude of the sum of the spins, $S^2$, was chosen as the first dynamical variable, while an angle describing how much the orbital angular momentum has precessed around the total angular momentum, $\phi_z$, was chosen as the second \cite{Chatziioannou:2017tdw,Kesden:2014sla}.  Another equivalent choice is to instead use the cosine of the angle between the orbital angular momentum and the total angular momentum, $\cos(\theta_L)$, as the first dynamical variable (keeping $\phi_z$ as the second).  Both choices have the disadvantage that their evolution is singular in the equal mass limit.  Klein proposed a different choice for the first dynamical quantity: $\delta \chi$, which is the difference of the projection of the reduced spins. With this choice, the evolution is well defined in the equal mass limit. For this reason this choice is preferred over using $S^2$ or $\cos(\theta_L)$ and $\phi_z$.  The relation between these variables and the angular momenta of the individual objects and orbital angular momentum is given in App.~\ref{app:variables} (see also Eqs.(10)-(11) in \cite{Klein:2021jtd}, where a detailed description of their evolution appears).\\
	
	Here, we extend previous results describing the precession of black holes to other compact objects by including the leading order finite-size effects through the quadrupole moment constants $\kappa_i$.  $\kappa_i$ is a coefficient that appears in the quadrupole moment of compact objects, $Q_i = -\kappa_i {\chi_i}^2 {m_i}^3$, and is determined by their properties, e.g. the equation of state.   For rotating black holes $\kappa$ is equal to one. For other compact objects, $\kappa \neq 1$. Specifically, for neutron stars, $\kappa > 1$ and it is larger for stiffer equations of state \cite{Poisson:1997ha}.\\
	
	For a system where $\kappa_i\neq1$, $\chi_{\rm eff}$ is no longer a conserved quantity.  This leaves us with a choice: either find a seventh conserved quantity to reduce the number of dynamical variables back to two, or work with three dynamical variables.  While the former option may seem simpler, as we will show in a later section, because of the more complicated form of the conserved quantity it will turn out to be easier to work with three dynamical variables.  Therefore, we use $\chi_{\rm eff}, \delta \chi$ and $\phi_z$ as our set of dynamical variables. As noted in \cite{Klein:2021jtd}, the equations for $\chi_{\text{eff}}$ and $\delta\chi$  take the form:
	\begin{equation}\label{eq:S002-SS001-EQ001}
		\begin{split}
			\left( \frac{d\delta\chi}{dt} \right)^2 = \frac{9 y^{11} }{4} A_{ \delta\chi }^2 \left( \delta\mu \delta\chi^3 + B\delta\chi^2 + C\delta\chi  + D   \right)
		\end{split}
	\end{equation}	
	and	
	\begin{equation}\label{eq:S002-SS001-EQ002}
		\begin{split}
			\left( \frac{d\chi_{\text{eff}} }{dt} \right)^2 = \frac{9 y^{11} }{4} A_{ \chi_{\text{eff}} }^2 \left( \delta\mu \delta\chi^3 + B\delta\chi^2 + C\delta\chi  + D   \right),
		\end{split}
	\end{equation}	
	where 	
	\begin{equation}\label{eq:S002-SS001-EQ003}
		\begin{split}
			A_{ \delta\chi } &= 1 + y \, A_{ \delta\chi , \delta\chi} \delta\chi + y \,  A_{ \delta\chi , \chi_{\text{eff}} } \chi_{\text{eff}},\\
			A_{ \chi_{\text{eff}}} &= y \,  A_{ \chi_{\text{eff}} , \delta\chi} \delta\chi + y \, A_{ \chi_{\text{eff}} , \chi_{\text{eff}} } \chi_{\text{eff}},\\
		\end{split}
	\end{equation}	
	with $y$ a post-Newtonian (PN) parameter related to the norm of the orbital angular momentum (in particular, $y=(M\omega)^\frac{1}{3}$ with $M$ the total mass, and $\omega$ the mean orbital frequency). The coefficients $B, C$, and $D$ depend on the conserved quantities, the PN parameter, and $\chi_{\rm eff}$, and are given in App.~\ref{app:variables} (see also \cite[App.~B]{Klein:2021jtd}). The coefficients in $A$ are 
	\begin{subequations}
		\begin{align}
			A_{\delta \chi, \delta \chi} & = \frac{\kappa_2-\kappa_1}{4},\\
			A_{\delta \chi, \chi_{\rm eff}} & = -\frac{\kappa_1 + \kappa_2 + 2}{4}, \\
			A_{\chi_{\rm eff}, \delta \chi} & = \frac{\kappa_1 + \kappa_2 - 2}{4}, \\
			A_{\chi_{\rm eff},\chi_{\rm eff}} & = \frac{\kappa_1-\kappa_2}{4}.
		\end{align} 
	\end{subequations}
	When $\kappa_i =1$, the equations reduce to those in \cite{Klein:2021jtd}.	
	To solve the dynamical equations, we will separate the behaviour of $\chi_{\text{eff}}$ and $\delta\chi$ in the next section. \\
	
	The amount the orbital angular momentum has precessed around the total angular momentum $\phi_z$ evolves according to	
	\begin{equation}\label{eq:S002-SS001-EQ005}
		\begin{split}
			\frac{d \phi_z }{dt} = \frac{1}{\sin^2(\theta_L)} \left[ \frac{d \hat{L}}{dt} \cdot \left( \hat{J} \times \hat{L} \right) \right] \; ,
		\end{split}
	\end{equation}
	where $\theta_L$ is the angle between the orbital angular momentum and the total angular momentum, and the hats on the orbital angular momentum $L$ and total angular momentum $J$ indicate that these are the direction vectors in the directions of the corresponding angular momentum.	
	While $\phi_z$ has a nice physical interpretation, it will also be useful to define a related angle $\zeta$, used in waveform generation \cite{Chatziioannou:2017tdw}
	\begin{equation}\label{eq:S002-SS001-EQ006}
		\begin{split}
			\frac{d \zeta }{dt} = - \cos(\theta_L)  \frac{d \phi_z }{dt} \; .
		\end{split}
	\end{equation}\\

	Once we include radiative effects, we need to take into account that the conserved quantities are no longer constant. The radiation reaction behavior is determined, largely, by two things, $dy/dt$ and $dJ/dt$, the latter of which is given by \cite{Klein:2021jtd}:	
	\begin{equation}\label{eq:S002-SS001-EQ005}
		\begin{split}
			\frac{d J }{dt} = - \frac{L}{2Jy} \frac{dy}{dt} \left( 2L + \chi_{\text{eff}} + \delta\mu \; \delta\chi \right),
		\end{split}
	\end{equation}
	where $\delta \mu$ is the difference of the reduced masses $\delta\mu = \mu_1 - \mu_2$.

	\section{Analytic Solution to the Precession Without Radiation Reaction}
	\label{sec:analytic-sol}	
	To solve the equations describing inspiralling precessing black hole or neutron star binaries on circular orbits, we take advantage of the fact that an (approximate) analytic solution exists for the conservative problem, as we will show in this section. In Sec. \ref{sec:radiation-reaction}, we include the effects of radiation reaction.
	To determine the conservative evolution, we separate the problem into a secular and a periodic part. The secular part of the solution is constant  for  $\delta\chi$, $\chi_{\text{eff}}$ and $J$ without considering radiation reaction. In contrast, the remaining key quantities, that is, $\phi_z$ and $\zeta$, have a secular part that evolves even in the absence of radiation reaction.  As such, for the former quantities, it is satisfactory to focus only on their periodic parts. The evolution of the angles $\phi_z$ and $\zeta$ requires additional treatment.  Therefore, we start by examining the periodic part of $\delta\chi$ and $\chi_{\text{eff}}$, before we discuss the more involved cases of $\phi_z$ and $\zeta$.\\
	
	As mentioned, in the black hole binary case (for which $\kappa_1 = \kappa_2 =1$), the effective spin  $\chi_{\text{eff}}$ is a constant of the conservative dynamics. This makes the process for solving the equation for $\delta\chi$ straightforward. In particular, the coefficients in the equation for $\delta \chi$ in Eq.~\eqref{eq:S002-SS001-EQ001} are all constant so that $\delta\chi$ oscillates between its minimum and maximum values, $\delta\chi_-$ and $\delta\chi_+$, respectively. The solution is then easily obtained by treating the solution as an average and oscillatory part, with $\delta \chi_-$ and $\delta \chi_+$ determining the average part of the solution as well as the amplitude of the oscillatory part. One is then only left to evolve the phase of the oscillatory part. \\
	
	If one or both objects in the binary are neutron stars so that $\kappa_i \neq 1$, $\chi_{\text{eff}}$ becomes dynamical and oscillates. As a result, the coefficients $B$, $C$, and $D$ appearing in Eq.~\eqref{eq:S002-SS001-EQ001} and Eq.~\eqref{eq:S002-SS001-EQ002} are now dynamical and so are its minima and maxima, $\delta \chi_-$ and $\delta \chi_+$.
	We overcome this complication by deriving an approximate linear relation between $\delta\chi$ and $\chi_{\text{eff}}$, so that $\chi_{\text{eff}}$ can be replaced by $\delta\chi$ in Eq.~\eqref{eq:S002-SS001-EQ001} and the resulting equation for $\delta \chi$ becomes independent of $\chi_{\text{eff}}$. Consequently, the roots on the right hand side of Eq.~\eqref{eq:S002-SS001-EQ001} again correspond to the true maximum and minimum of $\delta\chi$. This allows us to solve for the dynamics of $\delta \chi$. Then, using the linear relation between $\delta \chi$ and $\chi_{\rm eff}$, the evolution of $\chi_{\text{eff}}$ is trivially obtained.\\
	
	In the remainder of this section, we first derive this important linear relation in two different ways in Sec.~\ref{sec:approximate-relation-1} and \ref{sec:approximate-relation-2}. Next, we solve for the dynamics of $\delta \chi$ and $\chi_{\rm eff}$ in Sec.~\ref{sec:initdcandce}-\ref{sec:dpsi}. In Sec.~\ref{sec:consphizandzeta}, we discuss the dynamics for $\phi_z$ and $\zeta$.

	\subsection{Approximate relation between $\delta\chi$ and $\chi_{\text{eff}}$}
	\label{sec:approximate-relation-1}
	
	If the coefficients in Eq.~\eqref{eq:S002-SS001-EQ001} are constant, as is the case when both objects in the binary are black holes, the \emph{exact} solution to this equation is given by the Jacobi elliptic function
	\begin{equation}\label{eq:S003-SS001-EQ007}
		\delta \chi (t) = \delta \chi_- + \left( \delta \chi_+ - \delta \chi_-\right) \text{sn}^2(\psi'(t),m) \; ,
	\end{equation}
	where $\psi'$ describes the phase evolution of the precession and the parameter $m = \delta \mu \left(\delta \chi_+ - \delta \chi_-\right)/(\delta \chi_3 - \delta\mu \; \delta \chi_-)$ with $\delta \chi_3$ containing information of the largest root of the cubic equation on the right hand side of Eq.~\eqref{eq:S002-SS001-EQ001} (see Eq.~\eqref{eq:S003-SS005-EQ001}). 
	In the limit $m \to 0$, the Jacobi elliptic function reduces to the usual sine function. We will use this insight to approximate the solutions for $\delta \chi$ and $\chi_{\rm eff}$ in the generic case for which $\kappa_i\neq 1$ as 
	\begin{equation}\label{eq:S003-SS001-EQ001}
		\begin{split}
			\delta\chi &\approx \left< \delta\chi \right> + G_{\delta\chi} \sin(\psi),\\
			\chi_{\text{eff}} &\approx \left< \chi_{\text{eff}} \right> + G_{\chi_{\text{eff}} } \sin(\psi),\\
		\end{split}
	\end{equation}
	where 
	\begin{equation}\label{eq:S003-SS001-EQ002}
		\begin{split}
			\left< \delta\chi \right> &= \frac{1}{2}\left( \delta\chi_+ + \delta\chi_- \right), \\
			G_{\delta\chi} &= \frac{1}{2}\left( \delta\chi_+ - \delta\chi_- \right), \\
			\left< \chi_{\text{eff}} \right> &=  \frac{1}{2}\left( \chi_{\text{eff},+} +\chi_{\text{eff},-} \right), \\
			G_{\chi_{\text{eff}} } &= \frac{1}{2}\left( \chi_{\text{eff},+} - \chi_{\text{eff},-} \right).
		\end{split}
	\end{equation}
	
	Here we have used the brackets to indicate that there is some time average underlying these expressions, indeed the first expression corresponds to the actual precession average of $\delta\chi$ in Eq.~\eqref{eq:S003-SS001-EQ007} in the $m \to 0$ limit.  This precession average is performed on the precession timescale, $T_{pr} \sim O(y^{-5})$ which corresponds to the rate at which the precession phase, $\psi$, evolves.  This is in contrast to the radiation reaction timescale, $T_{rr} \sim O(y^{-8})$, on which the entire evolution occurs.  This separation of scales allows one to disregard the changes in these quantities during the precession averaging when adding radiation reaction.\\
	
	In these approximate expressions, we have use $\psi$ instead of $\psi'$ that appears in the original expressions, because we have simplified the $\sin^2$ term using the half angle formula and identified $2\psi'=\psi+\pi/2$ so that the final expression is expressible in terms of a sine function again.
	This simple form of the solution matches well with numerical evolutions of the coupled equations. Moreover, this form is very powerful as it will allow us to relate $\delta \chi$ and $\chi_{\rm eff}$. Note that due to the similarity in their derivatives in Eq.~\eqref{eq:S002-SS001-EQ001} and \eqref{eq:S002-SS001-EQ002}, when one solution has reached its extrema the other must have as well. This is the reason why both solutions oscillate with a single phase $\psi$.
	In analogy to the solutions in \cite{Chatziioannou:2017tdw, Klein:2021jtd}, we will refer to the ansatz in Eq.~\eqref{eq:S003-SS001-EQ001} as the ``m=0'' approximation. \\
	
	Operating within this approximation, Eq.~\eqref{eq:S002-SS001-EQ001} and \eqref{eq:S002-SS001-EQ002} can be related to obtain
	\begin{equation}\label{eq:S003-SS001-EQ003}
		\begin{split}
			A_{ \chi_{\text{eff}} }^2 \left( \frac{d\delta\chi}{dt} \right)^2 = A_{ \delta\chi }^2 \left( \frac{d\chi_{\text{eff}} }{dt} \right)^2.
		\end{split}
	\end{equation}
	Substituting the approximate definitions \eqref{eq:S003-SS001-EQ001} into the above relation then gives
	\begin{equation}\label{eq:S003-SS001-EQ004}
		( y A_{ \chi_{\text{eff}} , \delta\chi} \delta\chi + y A_{ \chi_{\text{eff}} , \chi_{\text{eff}} } \chi_{\text{eff}} )^2 G_{\delta\chi}^2  = (1 + y A_{ \delta\chi , \delta\chi} \delta\chi + y A_{ \delta\chi , \chi_{\text{eff}} } \chi_{\text{eff}})^2 G_{\chi_{\text{eff}} }^2 .
	\end{equation}	
	Evaluating this at $\psi=0$ relates the amplitude of one quantity, $G_{\chi_{\text{eff}} }$, to the amplitude of the other, $G_{\delta\chi}$\footnote{Note that the inverse of this expression is not well-defined in the black hole limit.
	}
	\begin{equation}\label{eq:S003-SS001-EQ005}
		\begin{split}
			\left| \frac{ G_{\chi_{\text{eff}} }  }{ G_{\delta\chi} } \right| = \left| \frac{ y A_{ \chi_{\text{eff}} , \delta\chi} \left< \delta\chi \right> + y A_{ \chi_{\text{eff}} , \chi_{\text{eff}} } \left< \chi_{\text{eff}} \right> }{ 1 + y A_{ \delta\chi , \delta\chi} \left< \delta\chi \right> + y A_{ \delta\chi , \chi_{\text{eff}} } \left< \chi_{\text{eff}} \right> }  \right| \; .
		\end{split}
	\end{equation}
	\\
	
	Next, solving for $\sin(\psi)$ in \eqref{eq:S003-SS001-EQ001}, we obtain the desired relation between $\chi_{\rm eff}$ and $\delta \chi$:	
	\begin{equation}\label{eq:S003-SS001-EQ006}
		\begin{split}
			\chi_{\text{eff}} = \left< \chi_{\text{eff}} \right> - \frac{ G_{\chi_{\text{eff}} }  }{ G_{\delta\chi} } \left< \delta\chi \right> + \frac{ G_{\chi_{\text{eff}} }  }{ G_{\delta\chi} } \delta\chi .
		\end{split}
	\end{equation}	
	Having already found the relation between the amplitudes $ G_{\chi_{\text{eff}} }  / G_{\delta\chi} $ in terms of the averages, this equation is entirely determined so long as the average values of $\chi_{\rm eff}$ and $\delta\chi$ are known.  Thus the desired result is achieved: an approximate linear relation in the absence of radiation reaction between $\delta\chi$ and $\chi_{\text{eff}}$ defined by their averages (which are known constants).	
	This relation will be used in Sec.~\ref{sec:dcandceamps} to obtain a new dynamical equation for $\delta \chi$ that no longer depends on the dynamics of $\chi_{\rm eff}$ but only on its average and amplitude.
	
	\subsection{Another perspective: the amplitude relation from a conserved quantity}
	\label{sec:approximate-relation-2}
	
	The relation between $\chi_{\rm eff}$ and $\delta \chi$ in \eqref{eq:S003-SS001-EQ006} holds approximately because it was derived assuming that the solutions for $\chi_{\rm eff}$ and $\delta \chi$ have the simple form given in Eq.~\eqref{eq:S003-SS001-EQ001}. 
	Here, we show that the amplitude relation in Eq.~\eqref{eq:S003-SS001-EQ005} holds more generically as long as the average/amplitude of $\chi_{\rm eff}$ and $\delta \chi$ can be understood purely as a sum/difference between the maxima and minima. \\
	
	Starting with \eqref{eq:S002-SS001-EQ001} and \eqref{eq:S002-SS001-EQ002}, the chain rule gives:
	\begin{equation}\label{eq:S003-SS002-EQ001}
		\begin{split}
			\left( \frac{d\chi_{\text{eff}} }{d \delta\chi } \right)^2 = \left( \frac{ y A_{ \chi_{\text{eff}} , \delta\chi}  \delta\chi  + y A_{ \chi_{\text{eff}} , \chi_{\text{eff}} }  \chi_{\text{eff}}  }{ 1 + y A_{ \delta\chi , \delta\chi} \delta\chi + y A_{ \delta\chi , \chi_{\text{eff}} } \chi_{\text{eff}} }  \right)^2 .
		\end{split}
	\end{equation}	
	Let us first discuss the special case with $ A_{ \chi_{\text{eff}} , \chi_{\text{eff}} } = 0 = A_{ \delta\chi , \delta\chi}$ for which one directly obtains a solution through its quadratures	
	\begin{equation}\label{eq:S003-SS002-EQ002}
		1 + y A_{ \delta\chi , \chi_{\text{eff}} } \chi_{\text{eff}}
		= \sqrt{ 1 + A_{ \chi_{\text{eff}} , \delta\chi}A_{ \delta\chi , \chi_{\text{eff}} } y^2 ( c_1 \delta\chi^2 + c_2 )  } ,
	\end{equation}
	where $c_1 = \pm 1$ arises from a choice of sign when taking the square root of \eqref{eq:S003-SS002-EQ001}, and $c_2$ is a constant of integration.  Squaring both sides gives
	\begin{equation}\label{eq:S003-SS002-EQ003}
		2y A_{ \delta\chi , \chi_{\text{eff}} } \chi_{\text{eff}} + y^2 A_{ \delta\chi , \chi_{\text{eff}} }^2  \chi_{\text{eff}}^2
		=  A_{ \chi_{\text{eff}} , \delta\chi}A_{ \delta\chi , \chi_{\text{eff}} } y^2 ( c_1 \delta\chi^2 + c_2 ) .
	\end{equation}
	While \eqref{eq:S003-SS001-EQ001} is an approximate relation for generic values of $\psi$, by definition of the average and amplitudes in  \eqref{eq:S003-SS001-EQ002}, Eq.~\eqref{eq:S003-SS001-EQ001} holds exactly when $\psi = \pm \pi/2$.  As a result, we can substitute Eq.~\eqref{eq:S003-SS001-EQ001} evaluated at $\psi=\pi/2$ into \eqref{eq:S003-SS002-EQ003} without loss of generality.  Subtracting off the same equation evaluated at $\psi=-\pi/2$ gives:
	\begin{equation}\label{eq:S003-SS002-EQ004}
		\begin{split}
			\left| \frac{ G_{\chi_{\text{eff}} }  }{ G_{\delta\chi} } \right| = \left| \frac{ y A_{ \chi_{\text{eff}} , \delta\chi} \left< \delta\chi \right>  }{ 1 + y A_{ \delta\chi , \chi_{\text{eff}} } \left< \chi_{\text{eff}} \right> } \right|,
		\end{split}
	\end{equation}
	where the factor of $c_1$ has been replaced in favor of expressing this as a relation in terms of the magnitudes of the amplitudes. This equation recovers the result in Eq.~\eqref{eq:S003-SS001-EQ005} for the case at hand: $ A_{\chi_{\text{eff}} , \chi_{\text{eff}} } = 0 = A_{ \delta\chi , \delta\chi}$. 
	
	The general case in which $ A_{ \chi_{\text{eff}} , \chi_{\text{eff}} } \neq 0 $ and $A_{ \delta\chi , \delta\chi}\neq 0$ is obtained in the same manner. Specifically, by making the substitution
	\begin{equation}\label{eq:S003-SS002-EQ005}
		\begin{split}
			Z &= \chi_{\text{eff}} - A_{ \chi_{\text{eff}} , \delta\chi} / y( A_{ \delta\chi , \delta\chi} A_{ \chi_{\text{eff}} , \chi_{\text{eff}} }  - A_{ \chi_{\text{eff}} , \delta\chi} A_{ \delta\chi , \chi_{\text{eff}} } )\\
			X &= \delta\chi + A_{ \chi_{\text{eff}} , \chi_{\text{eff}} } / y( A_{ \delta\chi , \delta\chi} A_{ \chi_{\text{eff}} , \chi_{\text{eff}} }  - A_{ \chi_{\text{eff}} , \delta\chi} A_{ \delta\chi , \chi_{\text{eff}} } ),
		\end{split}
	\end{equation}
	and using the chain rule, we find 	
	\begin{equation}\label{eq:S003-SS002-EQ006}
		\begin{split}
			\frac{dZ }{dX }  = \frac{ A_{ \chi_{\text{eff}} , \delta\chi} X + A_{ \chi_{\text{eff}} , \chi_{\text{eff}} } Z }{ A_{ \delta\chi , \delta\chi} X + A_{ \delta\chi , \chi_{\text{eff}} } Z }  .
		\end{split}
	\end{equation}
	The solution to this equation is implicitly given by 	
	\begin{equation}\label{eq:S003-SS002-EQ007}
		4 ( A_{ \delta\chi , \chi_{\text{eff}} } Z + A_{ \delta\chi , \delta\chi} X )^2 = - (\kappa_1 \kappa_2 - 1) X^2 + c_2,
	\end{equation}	
	where we have again denoted the constant of integration corresponding to the conserved quantity as $c_2$, to highlight that it arises in a manner similar as in Eq.~\eqref{eq:S003-SS002-EQ003}.  Again, evaluating this expression using \eqref{eq:S003-SS001-EQ001} at $\psi=\pi/2$ and subtracting off the same equation evaluated at $\psi=-\pi/2$, one finds the amplitude relation in Eq.~\eqref{eq:S003-SS001-EQ005}.\\
	
	As alluded to before, Eq.~\eqref{eq:S003-SS002-EQ007} points to a conserved quantity in the case with $\kappa_i\neq 1$.  In the black hole case, this conserved quantity was simply $\chi_{\text{eff}}$, but here it is nonlinear in $\delta\chi$ and $\chi_{\text{eff}}$.  Of course, in the limit $\kappa_i=1$, this equation simply states that $\chi_{\rm eff}$ is conserved.  In Eq.~\eqref{eq:S003-SS002-EQ003} this limit is slightly more obvious, $A_{ \chi_{\text{eff}} , \delta\chi}=0$ in this limit and (after absorbing other conserved quantities into our definition) we find that $\chi_{\rm eff}$ is conserved.\\
	
	While the existence of this conserved quantity would reduce the number of dynamic quantities from three to two, it is more useful to use the approximate linear relation instead.  This is because the linear relation maintains the cubic nature of Eq.~\eqref{eq:S002-SS001-EQ001} in terms of $\delta\chi$, meaning its roots can be found analytically. If instead the exact relation in Eq.~\eqref{eq:S003-SS002-EQ003} were used, it would make the roots of the equation too difficult to find analytically. Numerical root finding would make this method impractical computationally, for little benefit in accuracy.

	\subsection{Finding the initial averages of $\delta\chi$ and $\chi_{\text{eff}}$}
	\label{sec:initdcandce}
	
	In order to calculate the amplitudes of $\delta \chi$ and $\chi_{\rm eff}$, from Eqs.~\eqref{eq:S003-SS001-EQ005} and \eqref{eq:S003-SS001-EQ006}, it is clear that we need to know their average values. Consequently, the averages need to be calculated from known quantities (such as $J$, $y$ and the initial values of $\chi_{\text{eff}}$ and $\delta\chi$).  To do this, first we compute the numerical value of the first three derivatives for the initial values of $\chi_{\text{eff}}$ and $\delta\chi$, evaluated using Eq.~\eqref{eq:S002-SS001-EQ001}.  We then relate these values to the first three derivatives of the approximate relation Eq.~\eqref{eq:S003-SS001-EQ001}, given as:
	\begin{equation}\label{eq:S003-SS003-EQ001}
		\begin{split}
			\frac{d \delta\chi}{dt} &\approx  \frac{d\psi}{dt}G_{\delta\chi} \cos(\psi),\\
			\frac{d^2 \delta\chi}{dt^2} &\approx  -\left( \frac{d\psi}{dt} \right)^2 G_{\delta\chi} \sin(\psi),\\
			\frac{d^3 \delta\chi}{dt^3} &\approx  -\left( \frac{d\psi}{dt} \right)^3 G_{\delta\chi} \cos(\psi),
		\end{split}
	\end{equation}	
	where here we have not included the second and third derivatives of $\psi$ because they are zero in the absence of radiation reaction.  In the presence of radiation reaction they are not zero, but as a result of multi-scale analysis they correspond to the evolution of $\psi$ on the radiation reaction timescale and as a result their non-inclusion does not produce significant error.  We then solve the ensuing system of equations to obtain:	
	\begin{equation}\label{eq:S003-SS003-EQ002}
		\begin{split}
			\left< \delta\chi \right> &\approx \delta\chi - \delta\chi'' \delta\chi / \delta\chi''',\\
			\tan(\psi) &\approx \left| \left(\delta\chi''/\delta\chi'''\right)  \sqrt{ -\delta\chi'''/\delta\chi' } \right|,\\
			\left| \frac{d\psi}{dt} \right| &\approx \sqrt{-\delta\chi'''/\delta\chi'},\\
			\left| G_{\delta\chi} \right| &\approx  \sqrt{ (\delta\chi')^2/(d\psi/dt)^2 + (\delta\chi'')^2/(d\psi/dt)^4  },
		\end{split}
	\end{equation}	
	where the second equation is true up to a factor of $\pi$ and a plus or minus sign depending on the signs of the initial derivatives, to get around this we define the phase in such a way that the amplitude of $\delta\chi$ is always positive.  Since we are working in the conservative dynamics currently, the average, initial amplitude, and precession frequency are constant, when we include radiation reaction these will also evolve, the description of which is given later.
	
	\subsection{Finding the amplitudes of $\delta\chi$ and $\chi_{\text{eff}}$}
	\label{sec:dcandceamps}
	
	Now, with the averages in hand, the relation \eqref{eq:S003-SS001-EQ006} can be used to find the amplitudes.  To simplify the next set of calculations we define
	\begin{equation}\label{eq:S003-SS004-EQ001}
		\begin{split}
			\chi_{\text{eff}} = N_0 + N_1 \delta\chi,
		\end{split}
	\end{equation}
	where (as a result of \eqref{eq:S003-SS001-EQ006}) these coefficients are given by
	\begin{equation}\label{eq:S003-SS004-EQ002}
		\begin{split}
			N_0 &= \left< \chi_{\text{eff}} \right> - \frac{ G_{\chi_{\text{eff}} }  }{ G_{\delta\chi} } \left< \delta\chi \right>,\\
			N_1 &= \frac{ G_{\chi_{\text{eff}} }  }{ G_{\delta\chi} }.
		\end{split}
	\end{equation}
	Substituting this into \eqref{eq:S002-SS001-EQ001} gives a new equation of the form:
	\begin{equation}\label{eq:S003-SS004-EQ003}
		\begin{split}
			\left( \frac{d\delta\chi}{dt} \right)^2 = \frac{9 y^{11} }{4} A_{ \delta\chi }^2 \left( X_3 \delta\chi^3 + X_2 \delta\chi^2 + X_1 \delta\chi  + X_0   \right),
		\end{split}
	\end{equation}
	where	
	\begin{equation}\label{eq:S003-SS004-EQ004}
		\begin{split}
			X_0 &= D_3 N_0^3 + D_2 N_0^2 + D_1 N_0 + D_0,\\
			X_1 &= 3D_3 N_0^2 N_1 + C_2 N_0^2 + 2 D_2 N_0 N_1 + C_1 N_0 + D_1 N_1 + C_0,\\
			X_2 &= 3 D_3 N_0 N_1^2 + 2 C_2 N_0 N_1 + D_2 N_1^2 + B_1 N_0 + C_1 N_1 + B_0,\\
			X_3 &= D_3 N_1^3 + C_2 N_1^2 + B_1 N_1 +  \delta\mu.
		\end{split}
	\end{equation}
	In the above expression, we have organized the coefficients of $B, C$ and $D$ that appear in Eq.~\eqref{eq:S002-SS001-EQ001} in the following way:	
	\begin{equation}\label{eq:S002-SS001-EQ005}
		\begin{split}
			B &= B_0 + B_1 \chi_{\text{eff}},\\
			C &= C_0 + C_1 \chi_{\text{eff}} + C_2 \chi_{\text{eff}}^2 ,\\
			D &= D_0 + D_1 \chi_{\text{eff}} + D_2 \chi_{\text{eff}}^2 + D_3 \chi_{\text{eff}}^3 .
		\end{split}
	\end{equation}		
	The explicit form of these coefficients is given in App.~\ref{app:variables}.
	As discussed, because this cubic no longer implicates $\chi_{\text{eff}}$, the roots are now the true maximum and minimum values of $\delta\chi$, $\delta\chi_+$ and $\delta\chi_-$.  The amplitude of the oscillatory part of $\delta \chi$ is then obtained via its definition $G_{\delta\chi} = \frac{1}{2} ( \delta\chi_+ - \delta\chi_-)$.
	Finally, Eq.~\eqref{eq:S003-SS001-EQ006} can then be used to find the amplitude for $\chi_{\text{eff}}$.
	
	\subsection{Evolving the phase of $\delta \chi$ and $\chi_{\rm eff}$}
	\label{sec:dpsi}
	
	With both the averages and amplitudes of $\delta \chi$ and $\chi_{\rm eff}$ in hand, we are only left to determine the evolution of the phase $\psi$.  For reference we re-express the relevant derivative, Eq.~\eqref{eq:S003-SS004-EQ003}, in the most immediately useful form:
	\begin{equation}\label{eq:S003-SS005-EQ001}
		\begin{split}
			\left( \frac{d\delta\chi}{dt} \right)^2 = \frac{9 y^{11} }{4} X_3 A_{ \delta\chi }^2 \left( \delta\chi - \delta\chi_+ \right)\left( \delta\chi - \delta\chi_- \right)\left( \delta\chi - \frac{\delta\chi_3}{\delta\mu} \right) \; ,
		\end{split}
	\end{equation}
	where the roots of the cubic equation are ordered such that $\delta \chi_- \le \delta \chi_+ \le \delta \chi_3 / \delta \mu$.
	To get an expression for the phase evolution $\psi$, we substitute Eq.~\eqref{eq:S003-SS001-EQ001} into the equation above	and use the definition in Eq.~\eqref{eq:S003-SS001-EQ003} to simplify the expression. The resulting equation is
	\begin{equation}\label{eq:S003-SS005-EQ002}
		\begin{split}
			\left( \frac{d\psi}{dt} \right)^2 = \frac{9 y^{11} }{4} X_3 A_{ \delta\chi }^2 \left( \frac{\delta\chi_3}{\delta\mu}-  \left< \delta\chi \right> - G_{\delta\chi} \sin(\psi)   \right).
		\end{split}
	\end{equation}	
	This relation is not approximate, all of the nonlinearity in the original equation is accounted for. Instead of this exact result, it is preferable to use an averaged version of this equation so that the phase evolves at a fixed rate. Intuitively, the times  at which the phase would have evolved faster/slower than its average rate correspond to neglected higher order modes in its Fourier series.  (One could include these higher order modes in the definition in \eqref{eq:S003-SS001-EQ001} to account for this; we will not do that here.) To simplify the final expression of this average we first define	
	\begin{equation}\label{eq:S003-SS005-EQ003}
		\begin{split}
			\left< A_{ \delta\chi } \right> &=  1 + y A_{ \delta\chi , \delta\chi} \left<\delta\chi \right> + y A_{ \delta\chi , \chi_{\text{eff}} } \left< \chi_{\text{eff}} \right>, \\
			G_A &= y A_{ \delta\chi , \delta\chi} G_{\delta\chi } + y A_{ \delta\chi , \chi_{\text{eff}} } G_{ \chi_{\text{eff}} }.
		\end{split}
	\end{equation}	
	The average rate of change of $\psi$ is then given by	
	\begin{equation}\label{eq:S003-SS005-EQ004}
		\begin{split}
			\left( \frac{d\psi}{dt} \right)^2 \approx \frac{9 y^{11} }{8} X_3 ( \delta\chi_3/\delta\mu -  \left< \delta\chi \right> )( 2 \left< A_{ \delta\chi } \right>^2 + G_A^2 )  - G_{\delta\chi} G_A \left< A_{ \delta\chi } \right> ).
		\end{split}
	\end{equation}
	In the absence of radiation reaction, all quantities in the right hand side are constant and thus this can be integrated exactly to give:
	\begin{equation}\label{eq:S003-SS005-EQ005}
		\begin{split}
			\psi(t) \approx  \frac{3 y^{11/2} }{2^{3/2}} \sqrt{ X_3 ( \delta\chi_3/\delta\mu -  \left< \delta\chi \right> )( 2 \left< A_{ \delta\chi } \right>^2 + G_A^2 )  - G_{\delta\chi} G_A \left< A_{ \delta\chi } \right> ) } \;\; t + \psi(0).
		\end{split}
	\end{equation}
	
	\subsection{Evolving $\phi_z$ and $\zeta$}
	\label{sec:consphizandzeta}
	
	Given $\chi$ and $\chi_{\rm eff}$, we now turn to the slightly more complicated evolutions of $\phi_z$ and $\zeta$ as these variables have both an evolving periodic and an evolving secular part, also in the absence of radiation. The derivative of $\phi_z$ can be given approximately as (see appendix \ref{app:phiz}):	
	\begin{equation}\label{eq:eq:S003-SS006-EQ001}
		\begin{split}
			\frac{d\phi_z}{dt} \approx& \frac{J y^6}{2} \left\{ Q_4+Q_5\sin(\psi) + \frac{ H_0 + H_1 \sin(\psi) + H_2 \sin^2(\psi) + H_3 \sin^3(\psi) }{ ( 1 + H_- \sin(\psi))( 1 + H_+ \sin(\psi))} \right\}.
		\end{split}
	\end{equation}
	While this can be integrated exactly, it is useful to split $\phi_z$ into its secular and periodic part.  To do this, we rewrite the derivative of $\phi_z$ to separate terms with differing behavior:	
	\begin{equation}\label{eq:eq:S003-SS006-EQ002}
		\begin{split}
			\frac{d\phi_z}{dt} \approx& \frac{J y^6}{2} \left\{ Q_4 + \frac{H_2 H_+ H_- - H_3 H_- - H_3 H_+ }{{H_+}^2 {H_-}^2}  \right\} + \frac{J y^6}{2}\left\{ Q_5 + \frac{H_3}{H_+ H_-} \right\}\sin(\psi)\\
			& + \frac{J y^6}{2} \left\{ \frac{H_0 {H_+}^3 - H_1 {H_+}^2 + H_2 {H_+} - H_3}{( H_+ - H_-) {H_+}^2 ( 1 + H_+ \sin(\psi))} \right\} - \frac{J y^6}{2} \left\{ \frac{H_0 {H_-}^3 - H_1 {H_-}^2 + H_2 {H_-} - H_3}{( H_+ - H_-) {H_-}^2 ( 1 + H_- \sin(\psi))} \right\}.
		\end{split}
	\end{equation}	
	The first term is purely secular, the second term is purely periodic and the last two terms are mixed. To split the last two terms into secular and periodic parts, we precession average these terms. The precession average then contributes to the secular part, while the remaining part contributes to the periodic behavior. The resulting secular part is 	
	\begin{equation}\label{eq:eq:S003-SS006-EQ003}
		\begin{split}
			\frac{d \left< \phi_{z} \right> }{dt} \approx \Phi_0 + \Phi_+ + \Phi_-,
		\end{split}
	\end{equation}	
	and the periodic part is	
	\begin{equation}\label{eq:eq:S003-SS006-EQ004}
		\begin{split}
			\frac{d \phi_{z} }{dt} - \frac{d \left< \phi_{z} \right> }{dt} \approx \Phi_s \sin(\psi) + \Phi_+ \left( \frac{ \sqrt{1-{H_+}^2 } }{1 + H_+ \sin(\psi) } - 1 \right)  + \Phi_- \left( \frac{ \sqrt{1-{H_-}^2 } }{1 + H_- \sin(\psi) } - 1 \right) ,
		\end{split}
	\end{equation}	
	where $\Phi_0$, $\Phi_s$, $\Phi_+$, and $\Phi_-$ are independent of $\phi_z$ and $\zeta$ and defined in App.~\ref{app:phiz}.  The periodic part can be integrated exactly to yield
	\begin{equation}\label{eq:eq:S003-SS006-EQ005}
		\begin{split}
			\phi_{z} - \left< \phi_{z} \right> \approx& - \frac{\Phi_s}{ \dot{\psi} } \cos(\psi) + \frac{\Phi_+}{\dot{\psi} } \left( 2 \arctan\left( \frac{\tan(\psi/2) + H_+}{\sqrt{1 - {H_+}^2 } } \right) - \psi - \arcsin(H_+) \right)\\
			& + \frac{\Phi_-}{  \dot{\psi} }  \left(  2 \arctan\left( \frac{\tan(\psi/2) + H_-}{\sqrt{1 - {H_-}^2 } } \right) - \psi - \arcsin(H_-)  \right) .
		\end{split}
	\end{equation}
	The constant terms (involving arcsin) come from the constant of integration, and are used to ensure that the average of the periodic part is zero.

	The evolution for $\zeta$ is solved in a similar manner. Its derivative is given by $d \zeta/dt = - \cos(\theta_L) d\phi_z/dt$. Substituting Eq.~\eqref{eq:eq:S003-SS006-EQ002} and the expression for $\cos(\theta_L)$ gives
	\begin{equation}\label{eq:eq:S003-SS006-EQ006}
		\begin{split}
			\frac{d\zeta}{dt}  \approx - \left( \Theta_0 + \Theta_s \sin(\psi) \right)\left( \Phi_0 + \Phi_s \sin(\psi) +  \frac{ \Phi_+ \sqrt{1-{H_+}^2 } }{1 + H_+ \sin(\psi) }  + \frac{ \Phi_- \sqrt{1-{H_-}^2 } }{1 + H_- \sin(\psi) } \right)
		\end{split}
	\end{equation}	
	Splitting this into a secular and periodic part, we find that the secular part is
		\begin{equation}\label{eq:eq:S003-SS006-EQ007}
		\begin{split}
			\frac{d \left< \zeta \right> }{dt}  \approx - \Theta_0 ( \Phi_0 + \Phi_+ + \Phi_- ) - \frac{ \Theta_s \Phi_s}{2} - \frac{\Theta_s \Phi_+}{H_+}\left( \sqrt{1-{H_+}^2} - 1 \right) - \frac{\Theta_s \Phi_-}{H_-} \left( \sqrt{1-{H_-}^2 } - 1 \right),
		\end{split}
	\end{equation}
	and the periodic part is	
	\begin{equation}\label{eq:eq:S003-SS006-EQ008}
		\begin{split}
			\frac{d \zeta }{dt} - \frac{d \left< \zeta \right> }{dt}  \approx & - \Theta_0 \left( \frac{d \phi_{z} }{dt} - \frac{d \left< \phi_{z} \right> }{dt} \right) - \Theta_s \Phi_0 \sin(\psi) + \frac{ \Theta_s \Phi_s}{2} \cos(2\psi)\\
			& + \frac{\Theta_s \Phi_+}{H_+} \left( \frac{ \sqrt{1-{H_+}^2 } }{1 + H_+ \sin(\psi) } - 1 \right)  + \frac{\Theta_s \Phi_-}{H_+} \left( \frac{ \sqrt{1-{H_-}^2 } }{1 + H_- \sin(\psi) } - 1 \right).
		\end{split}
	\end{equation}	
	The periodic part can be integrated exactly to yield	
	\begin{equation}\label{eq:eq:S003-SS006-EQ009}
		\begin{split}
			\zeta-  \left< \zeta \right>  \approx & - \Theta_0 \left( \phi_{z} - \left< \phi_{z} \right> \right) - \frac{ \Theta_s \Phi_0 }{ \dot{\psi} } \cos(\psi) + \frac{ \Theta_s \Phi_s}{4 \dot{\psi} } \sin(2\psi)\\
			& + \frac{\Theta_s \Phi_+}{H_+} \left( 2 \arctan\left( \frac{\tan(\psi/2) + H_+}{\sqrt{1 - {H_+}^2 } } \right) - \psi - \arcsin(H_+) \right)  \\
			& + \frac{\Theta_s \Phi_-}{H_+} \left( 2 \arctan\left( \frac{\tan(\psi/2) + H_-}{\sqrt{1 - {H_-}^2 } } \right) - \psi - \arcsin(H_-) \right) .
		\end{split}
	\end{equation}
	
	\section{Adding Radiation Reaction}
	\label{sec:radiation-reaction}	
	
	We have discussed the complete conservative dynamics of spins in Sec.~\ref{sec:analytic-sol}. With gravitational radiation reaction included, the orbital frequency increases as a function of time, which can be obtained as an expansion in Post-Newtonian orders. The  total angular momentum $J$ and the average part of $\chi_{\rm eff}, \delta \chi$ become time-dependent, with corresponding evolution equations discussed below.  We do not discuss the evolution of the secular part of $\phi_z$ and $\zeta$ here as the previous expressions for their derivatives can simply be evaluated and evolved on the radiation reaction timescale, but with updated values for the averages of $J$, $\delta\chi$ and $\chi_{\text{eff}}$ at each step.
	
	\subsection{Evolving $J$}
	\label{J}
	
	Restating it here for simplicity, the radiation reaction equation for $J$ is
	\begin{equation}\label{eq:S004-SS001-EQ001}
		\begin{split}
			\frac{dJ}{dt} = \frac{-L}{2Jy} \frac{dy}{dt} \left( 2L + \delta\mu \delta\chi + \chi_{\text{eff}} \right).
		\end{split}
	\end{equation}	
	Similarly to before, this can be expressed in the ``m=0'' approximation, expanding the solution into a secular part and periodic part.  Here, the solution is slightly out of phase with $\delta\chi$/$\chi_{\text{eff}}$, so an additional term will be required to account for this.  Thus, the solution for $J$ should take the following form	
	\begin{equation}\label{eq:S004-SS001-EQ002}
		\begin{split}
			J \approx \left< J \right> + G_{J,s} \sin(\psi) + G_{J,c} \cos(\psi) 
		\end{split}
	\end{equation}	
	and we are left to determine $\left< J \right>, G_{J,s}$ and $G_{J,c}$.
	Because the secular part is much smaller than the periodic part, $G_{J,s}/\left< J \right>$ and $G_{J,c}/\left< J \right>$ are assumed to be small.  Substituting Eqs.~\eqref{eq:S003-SS001-EQ001} and \eqref{eq:S004-SS001-EQ002} into Eq.~\eqref{eq:S004-SS001-EQ001} and applying this approximation gives	
	\begin{equation}\label{eq:S004-SS001-EQ003}
		\begin{split}
			\frac{d \left< J\right>}{dt} + G_{J,s} \frac{d\psi}{dt} \cos(\psi) - G_{J,c} \frac{d\psi}{dt} \sin(\psi) &\approx \frac{-L}{2\left< J \right> y} \frac{dy}{dt}\left( 2L + \delta\mu \delta\chi + \chi_{\text{eff}} \right) \\
			&+ \frac{L}{2\left< J \right>^2 y}\frac{dy}{dt}\left( 2L + \delta\mu \delta\chi + \chi_{\text{eff}} \right) \left( G_{J,s} \sin(\psi) + G_{J,c} \cos(\psi) \right).
		\end{split}
	\end{equation}	
	$d \left< J\right>/dt$ is found by averaging both sides of this equation with respect to $\psi$ over one cycle:	
	\begin{equation}\label{eq:S004-SS001-EQ004}
		\begin{split}
			\frac{d \left< J\right>}{dt} =\frac{-L}{2\left< J \right> y} \frac{dy}{dt}\left( 2L + \delta\mu \left<\delta\chi\right> + \left<\chi_{\text{eff}}\right> \right) + \frac{L}{4 \left< J \right>^2 y}\frac{dy}{dt}G_{J,s} \left( \delta\mu G_{\delta\chi} + G_{\chi_{\text{eff}} } \right) ,
		\end{split}
	\end{equation}
	where we have treated $d\psi/dt$ as a constant for the purpose of the precession average, because it only evolves on the radiation reaction timescale.
	Another, alternative way of thinking about this is that the correction to this equation resulting from accounting for the change in $d\psi/dt$ will always be multiplied by one of the two amplitudes of $J$, this correction being small and the amplitudes of $J$ also being small means the resulting term is sufficiently small so that it can safely be neglected.  
	A system of equations for $G_{J,s}$ and $G_{J,c}$ can be obtained by multiplying by either $\sin(\psi)$ or $\cos(\psi)$ and averaging over $\psi$:	
	\begin{align}
		&G_{J,s} \frac{d\psi}{dt} = \frac{L}{2\left< J \right>^2 y}\frac{dy}{dt}G_{J,c} \left( 2L + \delta\mu \left<\delta\chi\right> + \left<\chi_{\text{eff}}\right> \right) \label{eq:S004-SS001-EQ005}
		\\
		&- G_{J,c} \frac{d\psi}{dt} = \frac{-L}{2\left< J \right> y} \frac{dy}{dt} \left( \delta\mu G_{\delta\chi} + G_{\chi_{\text{eff}} } \right) + \frac{L}{2\left< J \right>^2 y}\frac{dy}{dt} G_{J,s} \left( 2L + \delta\mu \left<\delta\chi\right> + \left<\chi_{\text{eff}}\right> \right).\label{eq:S004-SS001-EQ006}
	\end{align}	
	Eqs. \eqref{eq:S004-SS001-EQ004}-\eqref{eq:S004-SS001-EQ006} form a system of equations for the desired unknowns: $d\left< J \right>/dt$, $G_{J,s}$, and $G_{J,c}$.  In previous works \cite{Chatziioannou:2017tdw, Klein:2021jtd} the value of $J$ is then used to calculate the roots of $\delta\chi$, instead here we use this derivative of the average of $J$ to find the derivatives of the averages of $\delta\chi$ and $\chi_{\text{eff}}$.
	
	\subsection{Evolving $\left< \delta\chi \right>$ and $\left< \chi_{\text{eff}} \right>$}
	\label{ddcandce}
	
	After using the results of Sec.~\ref{sec:initdcandce} we can obtain the averages $\left< \delta\chi \right>$ and $\left< \chi_{\text{eff}} \right>$ at the initial time. In the presence of radiation reaction we must then evolve these averages.  First we evolve the roots ($\delta\chi_+$, $\delta\chi_-$, $\chi_{\text{eff},+}$, $\chi_{\text{eff},-}$) and then use the relation between these roots and the averages (Eq.~\eqref{eq:S003-SS001-EQ002} to evolve the averages.  To find these derivatives, we start with their definitions as the roots of the following equations:
	
	\begin{align}
		\label{eq:S004-SS002-EQ001}
		&\delta\mu {\delta\chi_+}^3 + (B_0 + B_1 \chi_{\text{eff},+} ) {\delta\chi_+}^2 + (C_0 + C_1 \chi_{\text{eff},+} + C_2 {\chi_{\text{eff},+}}^2 ) {\delta\chi_+} \notag \\
		  &+ (D_0 + D_1 \chi_{\text{eff},+} + D_2 {\chi_{\text{eff},+}}^2 + D_3 {\chi_{\text{eff},+}}^3 ) = 0, \\\notag\\
		\label{eq:S004-SS002-EQ002}
		&\delta\mu {\delta\chi_-}^3 + (B_0 + B_1 \chi_{\text{eff},-} ) {\delta\chi_-}^2 + (C_0 + C_1 \chi_{\text{eff},-} + C_2 {\chi_{\text{eff},-}}^2 ) {\delta\chi_-} \notag \\
		  &+ (D_0 + D_1 \chi_{\text{eff},-} + D_2 {\chi_{\text{eff},-}}^2 + D_3 {\chi_{\text{eff},-}}^3 ) = 0, .
	\end{align}
	
	Using the implicit function theorem we can get two equations, one satisfied by the positive roots $\delta\chi_+$ and $\chi_{\text{eff},+}$ (the derivative of Eq.~\eqref{eq:S004-SS002-EQ001}), and one satisfied by the negative roots $\delta\chi_-$ and $\chi_{\text{eff},-}$ (the derivative of Eq.~\eqref{eq:S004-SS002-EQ002}):	
	\begin{align}
		\label{eq:S004-SS002-EQ003}
		&B_0' \delta\chi_+^2 + \left( C_0' + C_1'\chi_{\text{eff},+} \right) \delta\chi_+ + \left( D_0' + D_1'\chi_{\text{eff},+} + D_2'\chi_{\text{eff},+} \right) 
		+ \left( 3\delta\mu \delta\chi_+^2 + 2 B \delta\chi_+ + C \delta\chi_+ \right) \delta\chi_+' \notag\\
		&+\left( B_1\chi_{\text{eff},+} \delta\chi_+^2+ (C_1 + 2C_2 \chi_{\text{eff},+})\delta\chi_+ + ( D_1 + 2D_2 \chi_{\text{eff},+}+ 3D_3 \chi_{\text{eff},+}^2 )  \right) \chi_{\text{eff},+}' = 0, \\\notag\\
		\label{eq:S004-SS002-EQ004}
		&B_0' \delta\chi_-^2 + \left( C_0' + C_1'\chi_{\text{eff},-} \right) \delta\chi_- + \left( D_0' + D_1'\chi_{\text{eff},-} + D_2'\chi_{\text{eff},-} \right) 
		+ \left( 3\delta\mu \delta\chi_-^2 + 2 B \delta\chi_- + C \delta\chi_- \right) \delta\chi_-' \notag \\
		&+\left( B_1\chi_{\text{eff},-} \delta\chi_-^2+ (C_1 + 2C_2 \chi_{\text{eff},-})\delta\chi_- + ( D_1 + 2D_2 \chi_{\text{eff},-}+ 3D_3 \chi_{\text{eff},-}^2 )  \right) \chi_{\text{eff},-}' = 0 ,
	\end{align}	
	where the primes denote time derivatives.  These can be obtained using the definitions of the coefficients, $B$, $C$, and $D$ in conjunction with the time derivatives of $J$ and $y$, as an example $B_0'$ is given as	
	\begin{equation}\label{eq:S004-SS002-EQ008}
		\begin{split}
			B_0' = \frac{\partial B_0}{\partial J} \frac{d \left< J \right>}{dt} + \frac{\partial B_0}{\partial y} \frac{d y}{dt}
		\end{split}
	\end{equation}	
	
	A second set of relations follow from the derivatives directly, giving a complete set of equations:
	\begin{align}
		\label{eq:S004-SS002-EQ005}
		\chi_{\text{eff},+}' &= \frac{ y A_{ \chi_{\text{eff}} , \delta\chi} \delta\chi_+ + y A_{ \chi_{\text{eff}} , \chi_{\text{eff}} } \chi_{\text{eff},+} }{ 1 + y A_{ \delta\chi , \delta\chi} \delta\chi_+ + y A_{ \delta\chi , \chi_{\text{eff}} } \chi_{\text{eff},+} } \delta\chi_+', \\
		\label{eq:S004-SS002-EQ006}
		\chi_{\text{eff},-}' &= \frac{ y A_{ \chi_{\text{eff}} , \delta\chi} \delta\chi_- + y A_{ \chi_{\text{eff}} , \chi_{\text{eff}} } \chi_{\text{eff},-} }{ 1 + y A_{ \delta\chi , \delta\chi} \delta\chi_- + y A_{ \delta\chi , \chi_{\text{eff}} } \chi_{\text{eff},-} } \delta\chi_-'.
	\end{align}
	The derivative of the averages are then straightforwardly obtained from the solutions to these via:
	\begin{subequations}\label{eq:S004-SS002-EQ007}
		\begin{align}			
			\frac{d \left< \delta\chi \right> }{dt} &= \frac{1}{2}\left( \delta\chi_+' + \delta\chi_-' \right)\\
			\frac{d \left< \chi_{\text{eff}} \right> }{dt}&= \frac{1}{2}\left( \chi_{\text{eff},+}' + \chi_{\text{eff},-}' \right)
		\end{align}
	\end{subequations}
	
	In the absence of radiation reaction the derivatives of $y$ and $J$ are zero and consequently the derivatives of the coefficients ($B_0'$, $B_1'$, $C_0'$, etc) are zero.  The solution to the resulting system of equations is $\delta\chi_+'=\delta\chi_-'=\chi_{\text{eff},+}'=\chi_{\text{eff},-}'=0$. Thus, we nicely recover that in the absence of radiation reaction the averages do not evolve.  
	
	\subsection{Summarizing}
	
\begin{figure}
		\begin{tikzpicture}[node distance = 2cm and 4cm]
			\node (start) [startstop] {Start};
			
			\node (init1) [init, right=0.5cm of start] { Initialize $\left< \delta\chi \right>(0)$, $\left< \chi_{\text{eff}} \right>(0)$, and $\psi(0)$  (Section \ref{sec:initdcandce}) };
			\draw [arrow] (start) -- (init1);
			
			\node (init2) [init, right=0.5cm of init1] { Initialize $\left< J\right>$(0) (Section \ref{J}) };
			\draw [arrow] (init1) -- (init2);
			
			\node (init3) [init, right=0.5cm of init2] { Initialize $\left< \phi_z\right>(0)$  and $\left< \zeta\right>$(0) (Section \ref{sec:consphizandzeta}) };
			\draw [arrow] (init2) -- (init3);
			
			\node (avgevo1) [evo, below=2cm of init3] { Evolve $\psi$ (Section \ref{sec:dpsi}) };
			\draw [arrow] (init3) -- (avgevo1);
			
			\node (avgevo2) [evo, below left=2cm and 0.5cm of init3] { Evolve $\left< J\right>$ (Section \ref{J}) };
			\draw [arrow] (avgevo1) -- (avgevo2);
			
			\node (avgevo3) [evo, below left=2cm and 4.5cm of init3] { Evolve $\left< \delta\chi \right>$ and $\left< \delta\chi \right>$ (Section \ref{ddcandce}) };
			\draw [arrow] (avgevo2) -- (avgevo3);
			
			\node (avgevo4) [evo, below left=2cm and 8cm of init3] { Evolve $\left< \phi_z \right>$ and $\left< \zeta \right>$ (Section \ref{sec:consphizandzeta}) };
			\draw [arrow] (avgevo3) -- (avgevo4);
			
			\node (amp1) [amp, below=2cm of avgevo4] { Get periodic part of $\delta\chi$ and $ \chi_{\text{eff}} $ (Section \ref{sec:dcandceamps}) };
			\draw [arrow] (avgevo4) -- (amp1);
			
			\node (amp2) [amp, right=0.5cm of amp1] { Get periodic part of $J$ (Section \ref{J}) };
			\draw [arrow] (amp1) -- (amp2);
			
			\node (amp3) [amp, right=0.5cm of amp2] { Get periodic part of $\phi_z$ and $\zeta$ (Section \ref{sec:consphizandzeta}) };
			\draw [arrow] (amp2) -- (amp3);
			
			\node (dec) [deci, below=1cm of avgevo1] { Are we finished? };
			\draw [arrow] (amp3) -- (dec);
			
			\draw [arrow] (dec) -- node[anchor=east] {no} (avgevo1);
			
			\node (end) [startstop, below=1cm of dec] {end };
			\draw [arrow] (dec) -- node[anchor=east] {yes} (end);
			\end{tikzpicture}
		\caption{This flow chart summarizes the key steps in solving for the precession equations of compact objects (with possibly $\kappa \neq 1$) on circular orbits. The parts in red initialize the averages, the parts in yellow update the averages, and the parts in green evolve the periodic parts of the solutions. }
		\label{fig:flow-chart}
	\end{figure}
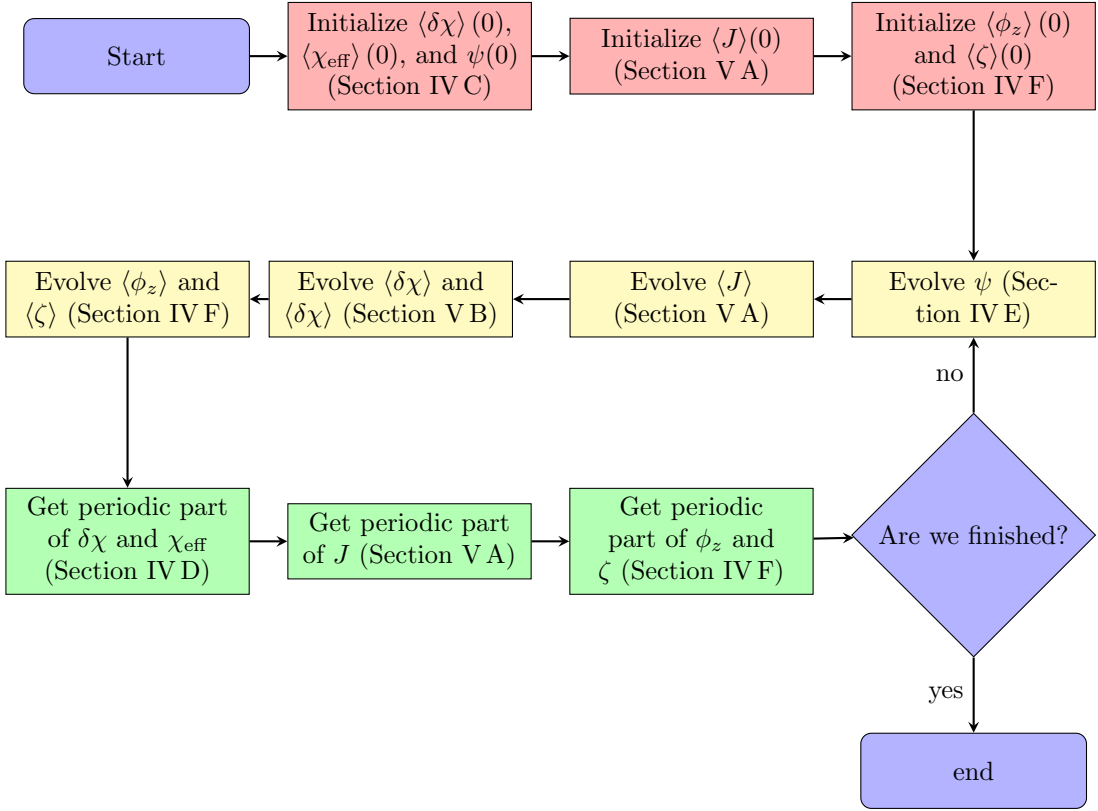

	At this point, we have discussed all the equation necessary for describing the spin dynamics under the influence of radiation reaction. Our method is summarized in Fig.~\ref{fig:flow-chart} in the form of a flow chart. The general idea can be broken up into several blocks, which can be described generally as: initializing the averages (red), updating the averages (yellow), and calculating the periodic part (green).  
	The first two deal with the secular parts, while the latter is critical for accounting for the effects of precession.  
	Once the evolution of these key quantities are known, the waveform is generated using standard techniques (see e.g. \cite[Sec.~IV]{Klein:2021jtd}).  Following \cite{Chatziioannou:2017tdw}, we refer to the resulting wavefor as a frequency domain waveform because, while the equations are evolved in the time domain, these time domain solutions can be transformed directly to frequency domain waveforms via the method of shifted uniform asymptotics (as opposed to constructing a time domain waveform then transforming this to the frequency domain).  We do not present the equations to produce this waveform here because they are identical to those found in \cite[Sec.~IV]{Klein:2021jtd}.  In the analysis here, we only use the leading term in $dy/dt = (32\eta/5) y^9$ unless otherwise specified. When constructing realistic waveforms for actual analysis of gravitational waves, this should be updated with the most current PN expression for $dy/dt$.  Thus, to include the effects of higher PN results for the nonspinning part one should adjust $dy/dt$ as well as modify the expressions for the nonspinning part appearing in the waveform generation.  
	
	\section{Comparing models}
\label{sec:comparison}
In this section, we first establish the validity of our method in Sec.~\ref{sec:effectiveness} before we compare waveforms of various binaries in Sec.~\ref{sec:effect-kappa}. In Sec.~\ref{sec:effectiveness} we compute the mismatch between the waveform constructed according to the discussion in Sec.~\ref{sec:analytic-sol},\ref{sec:radiation-reaction} and the waveform following numerical evolution of spins.  The latter section describe the key results of this paper: its shows that the quadrupole $\kappa_i$ has an observable imprint on the waveform, at least for some of the spin/orbital configurations.

\subsection{Model validation}
\label{sec:effectiveness}	
To evaluate the effectiveness of the proposed method, we compare the method to the fully numerically evolved precession equations Eq.~\eqref{eq:eq:S00B-SS001-EQ008}.  Our method is better than simply numerically evolving the precession equations directly because of the timescales on which they can be accurately evolved (while still accounting for the full effects of precession).  To numerically evolve the precession equations one must evolve them on the precession timescale, otherwise they run the risk of diverging.  These divergences usually occur as a result of one point lying outside the upper or lower bound on $\delta\chi$ or $ \chi_{\text{eff}} $, the solution then diverges secularly.  To avoid this behavior, one typically has to use an overly expensive number of points in the evolution to keep the error low. Our solution has no such issues. Since we are evolving the average values, there is no chance of such an issue happening in the first place and it can be accurately evolved on the radiation reaction timescale.  Because of the different timescales on which the system must be evolved, our method takes on average 5.1ms evolving from 10-100Hz with a single CPU for the system configurations in this section, while evolving the precession equations numerically (while keeping enough points to avoid the aforementioned divergence) takes on average 54ms evolving from 10-100Hz with a single CPU for the same configurations.  For high SNR events with better detector sensitivity, the waveform will be longer so that this speed-up factor will be larger. Our method is only constrained by how many frequency points one wants to evaluate, while the full numerical evolution is constrained by the precession timescale.\\

As concrete examples for comparison, we examine two systems: a binary comprised of two neutron stars and a binary consisting of one neutron star and one black hole.  For these comparisons we will use $\chi_1 = 0.4$ and $\chi_2 = 0.7$ for the spins, we set the angles of $\vec{s}_1$ and $\vec{s}_2$ relative to $\hat{L}$ as $\theta_{s1} = \pi/20$, $\phi_{s1} = 0$, $\theta_{s2} = \pi/4$, $\phi_{s2} = \pi/10$.  The only difference between the NSNS and NSBH systems we analyze are the masses and spin-induced quadrupole moment constants: 
\begin{itemize}
	\item For NSNS binaries, we will use $M_1 = 2.6M_\odot, M_2 = 1.5M_\odot,  \kappa_1 = 2.5$, $\kappa_2 = 3.5$.
	\item For NSBH binaries, we will use $M_1 = 23.0M_\odot, M_2 = 2.6M_\odot,  \kappa_1 = 1.0$ and $\kappa_2 = 2.5$.
\end{itemize} 
Here we have chosen values of $\kappa$ that are smaller than the typical range for neutron stars ($\kappa \sim 4-8$ \cite{Steinhoff:2021dsn}) to be conservative in examining whether this is potentially measurable. To facilitate the comparison, the tidal effects of NSs are not included here, although their implementation is straightforward to add.\\

To validate the effectiveness of our method, we start by looking at several spin variables as a function of frequency to gain insight into the error resulting from the approximations we made 
to see how we can improve the accuracy in the future.  To facilitate this, we separate the error into secular and periodic errors.  We start with $\delta\chi$ and $\chi_{\text{eff}}$, since they form two of our three dynamical variables, then we examine the variables more relevant to waveform generation: $\phi_z$ and $\theta_L$ (we exclude $\zeta$ because its evolution is similar to that of $\phi_z$).  Finally we randomize spin configurations and compute the fidelity/mismatch for various NSBH/NSNS systems, using the mismatch as a measure of the accuracy of our spin evolution scheme.  
\\

In the logarithmic error plots below, one can easily determine whether the secular evolution or the periodic evolution dominates the error budget. The presence of a series of sharp dips means that the error approaches zero repeatedly, implying that the periodic part is the main source of error.  When these sharp dips are absent, the secular error dominates. When there is a transition from the periodic part to the secular part being the dominant source of error, every other peak will gradually decrease in magnitude until there are no more sharp dips. This is because two consecutive peaks correspond to periods where the quantity is below/above what it should be.  As an example, if the secular error diverges to positive values, the quantity will gradually spend less time being ``too negative'' and more time being ``too positive'', until it is never ``too negative'', as a result the peaks that correspond to being ``too negative'' will diminish in size until they vanish altogether.\\

\begin{figure}[H]
	\centering
	\begin{subfigure}{.5\textwidth}
		\centering
		\includegraphics[width=0.9\textwidth]{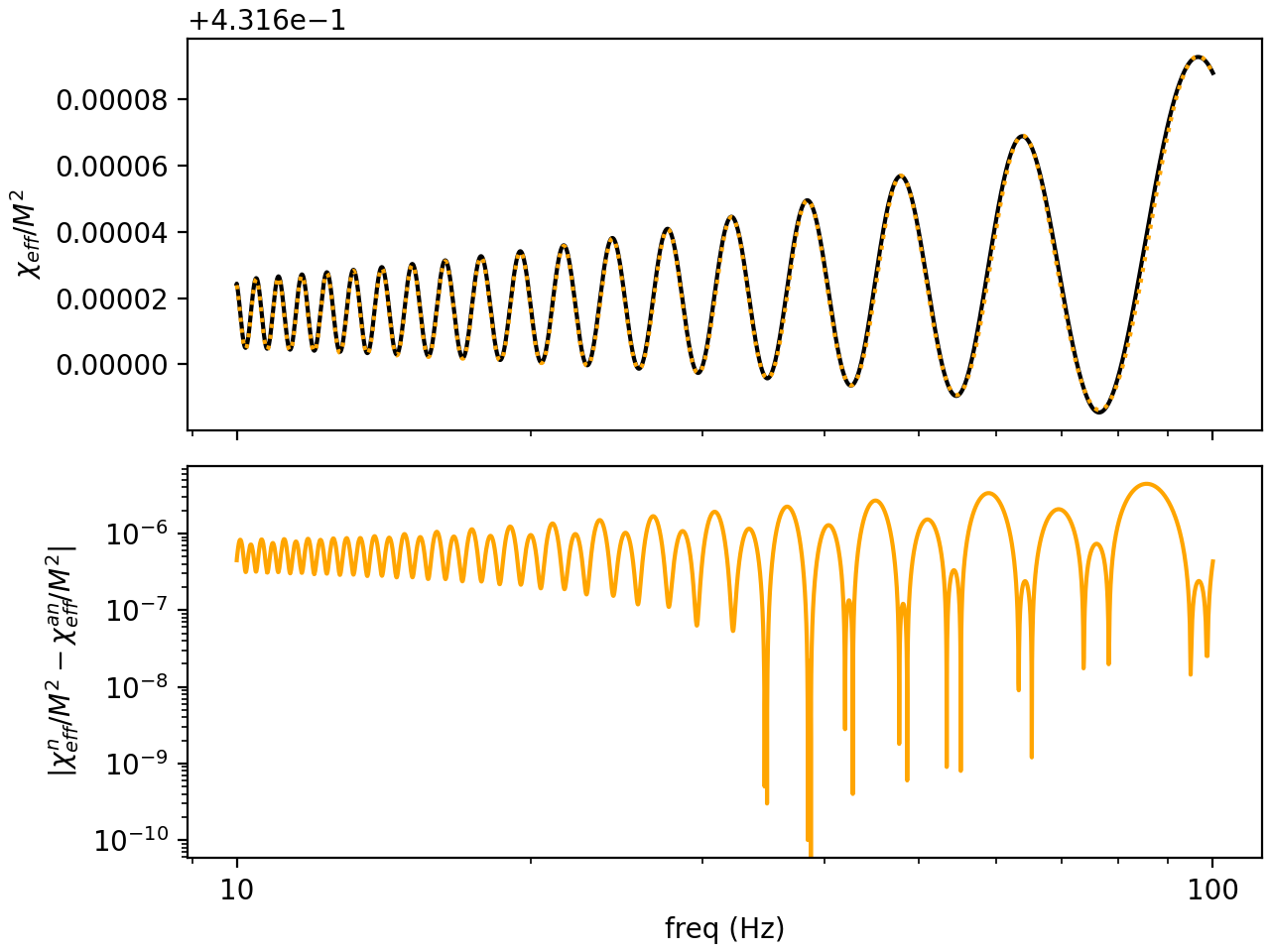}
		\caption[short]{NSNS binary}
	\end{subfigure}%
	\begin{subfigure}{.5\textwidth}
		\centering
		\includegraphics[width=0.9\textwidth]{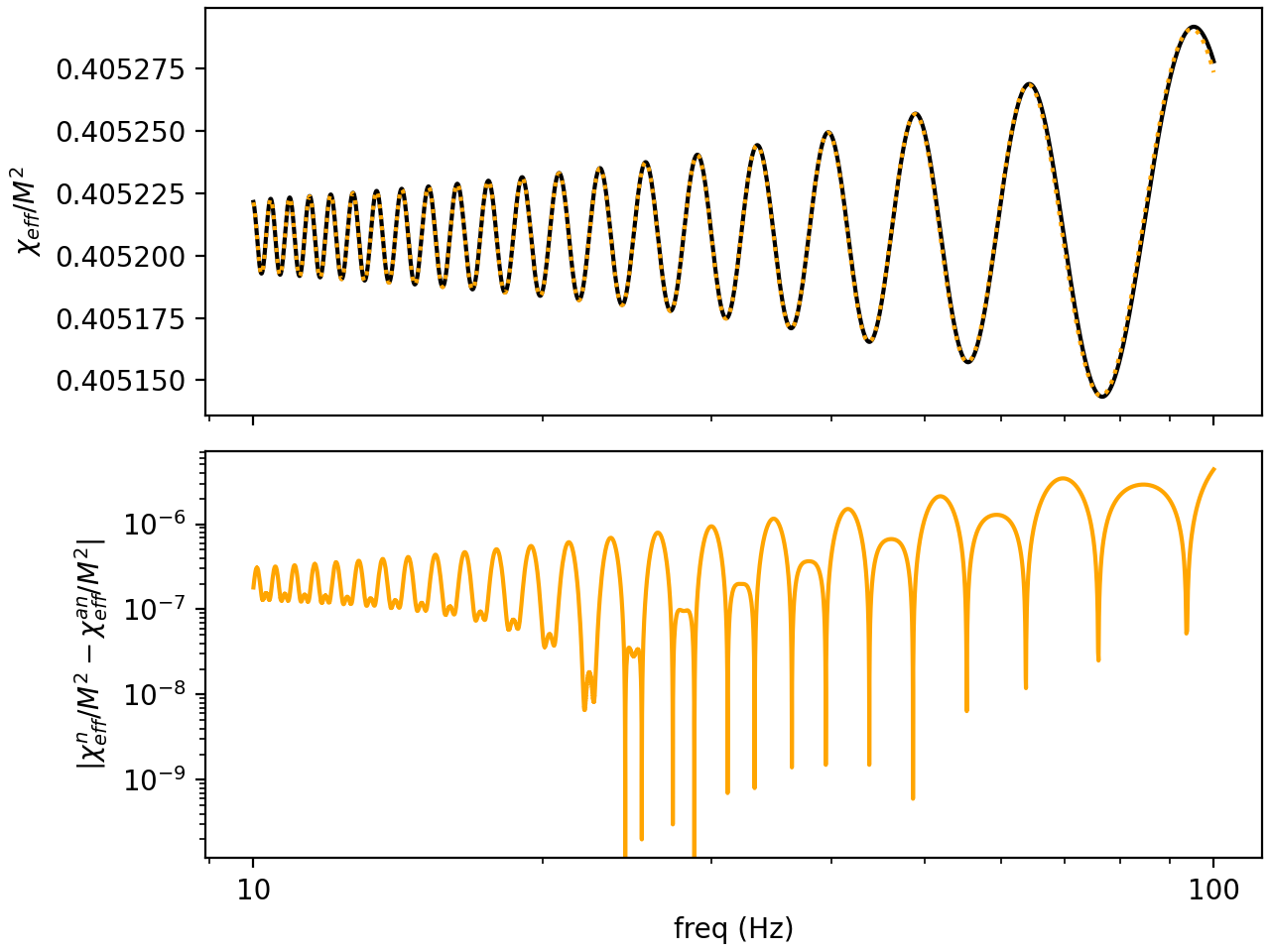}
		\caption[short]{NSBH binary}
	\end{subfigure}
	\begin{subfigure}{.5\textwidth}
		\centering
		\includegraphics[width=0.9\textwidth]{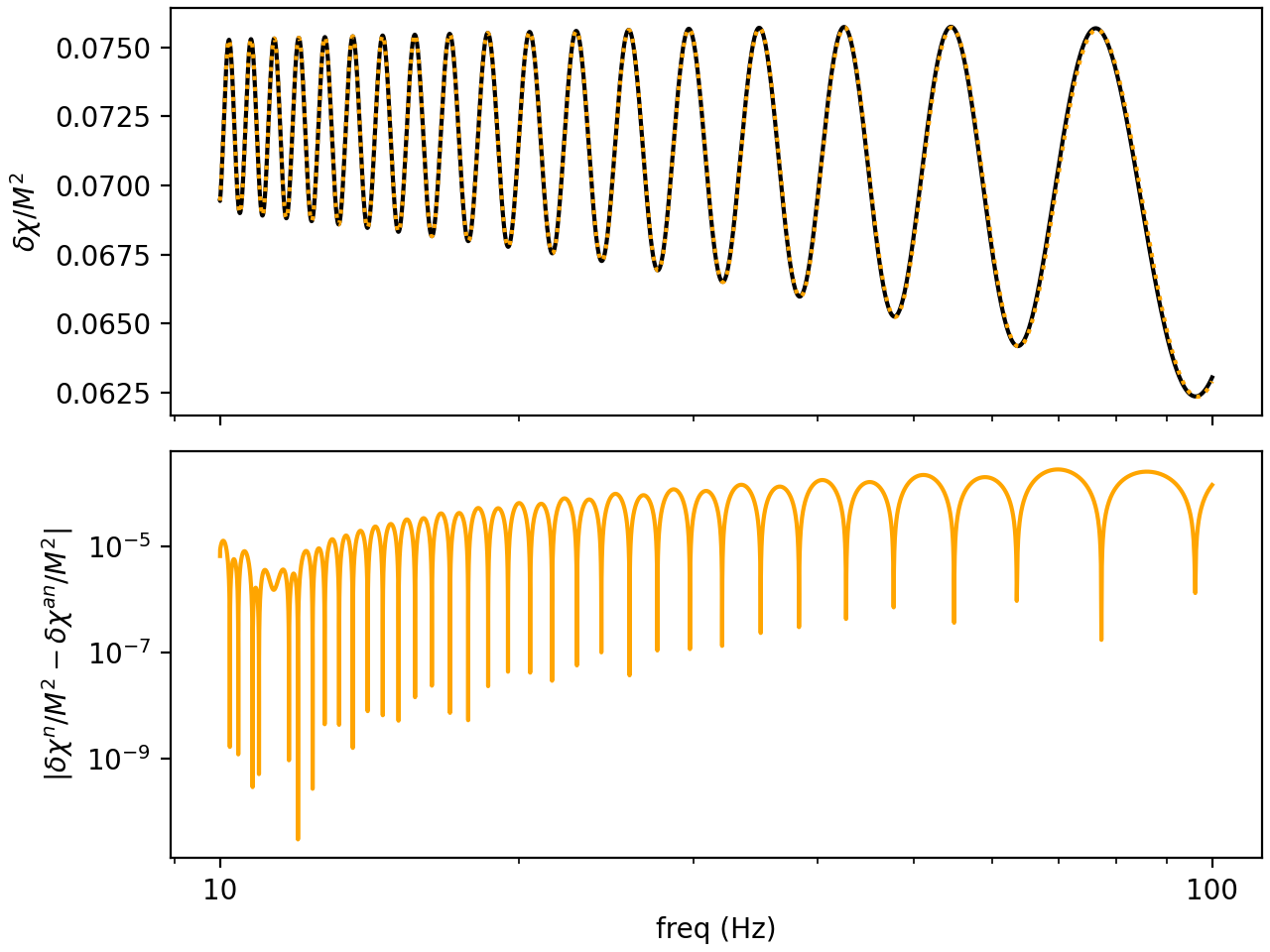}
		\caption[short]{NSNS binary}
	\end{subfigure}%
	\begin{subfigure}{.5\textwidth}
		\centering
		\includegraphics[width=0.9\textwidth]{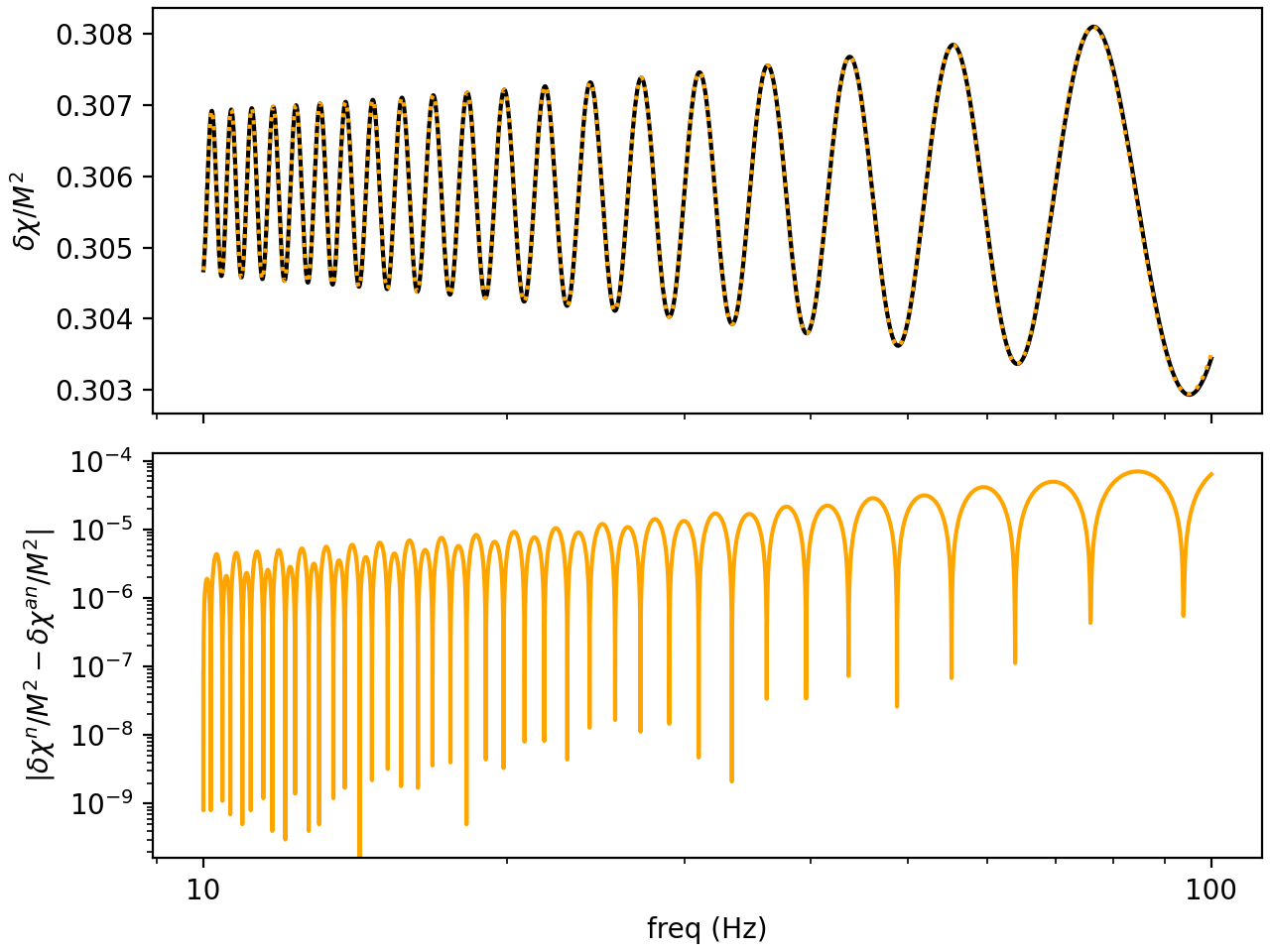}
		\caption[short]{NSBH binary}
	\end{subfigure}%
	\caption[short]{Plot of $\chi_{\text{eff}}/M^2$ vs frequency for the NSNS (NSBH) binary in panel a (b), and $\delta\chi/M^2$ for the NSNS (NSBH) binary in panel c (d).  In each plot the black curve is the numerically evolved system, while the orange curves are the solution found using the method outlined in this paper. The second plot in each panel is the absolute error. }
	\label{fig:numerical-vs-ourmethod-chis}
\end{figure}

We begin by comparing the numerical evolution of $\delta\chi$ and $\chi_{\text{eff}}$ to our semi-analytic approach in Fig.~\ref{fig:numerical-vs-ourmethod-chis}. First, note that there is good agreement between the precession frequency of the numerically evolved spins and the spins evolved using the methods in this paper. The error is largest for $\chi_{\rm eff}$ in the NSNS binary when the frequencies are high, but in most cases, the difference is not visible. Obtaining an accurate precession frequency is critical in reducing the error of the final solution, because small phase errors produce secular errors of the same order in the oscillation amplitude, which can introduce large periodic errors in the final result. 	
Second, while the agreement between the numerical and our method is good at the extrema of the oscillations, there is some disagreement in between.  This error is a result of approximating the solutions by their first mode only (i.e., the ``m=0'' approximation). Finally, there is a slight error in the amplitude and average of the oscillations. These latter error sources are the main contributions to the error in the final results. To see this, we start by examining $\phi_z$. 

\begin{figure}[H]
	\centering
	\begin{subfigure}{.5\textwidth}
		\centering
		\includegraphics[width=0.9\textwidth]{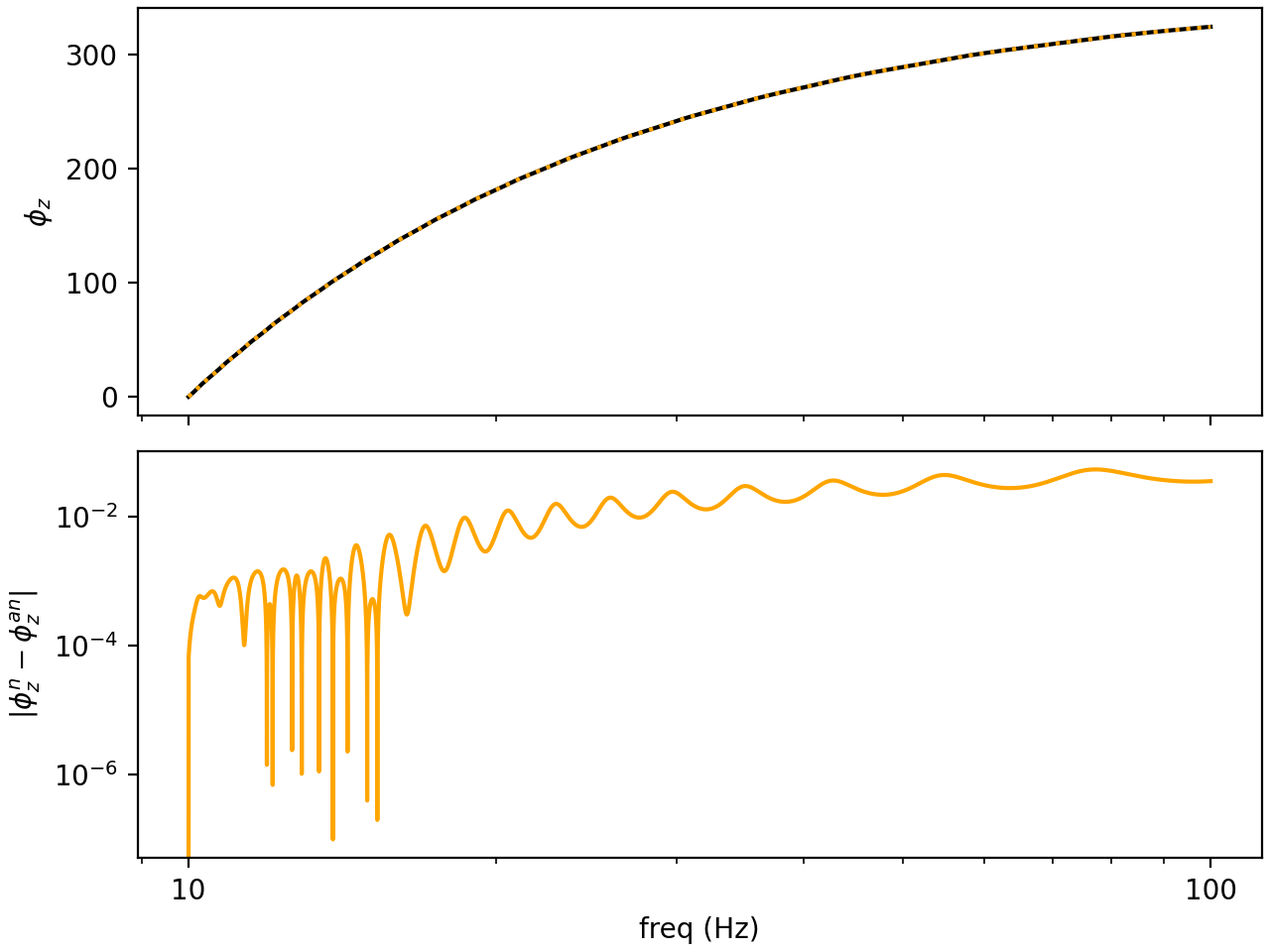}
		\caption[short]{NSNS Binary $\phi_z$}
	\end{subfigure}%
	\begin{subfigure}{.5\textwidth}
		\centering
		\includegraphics[width=0.9\textwidth]{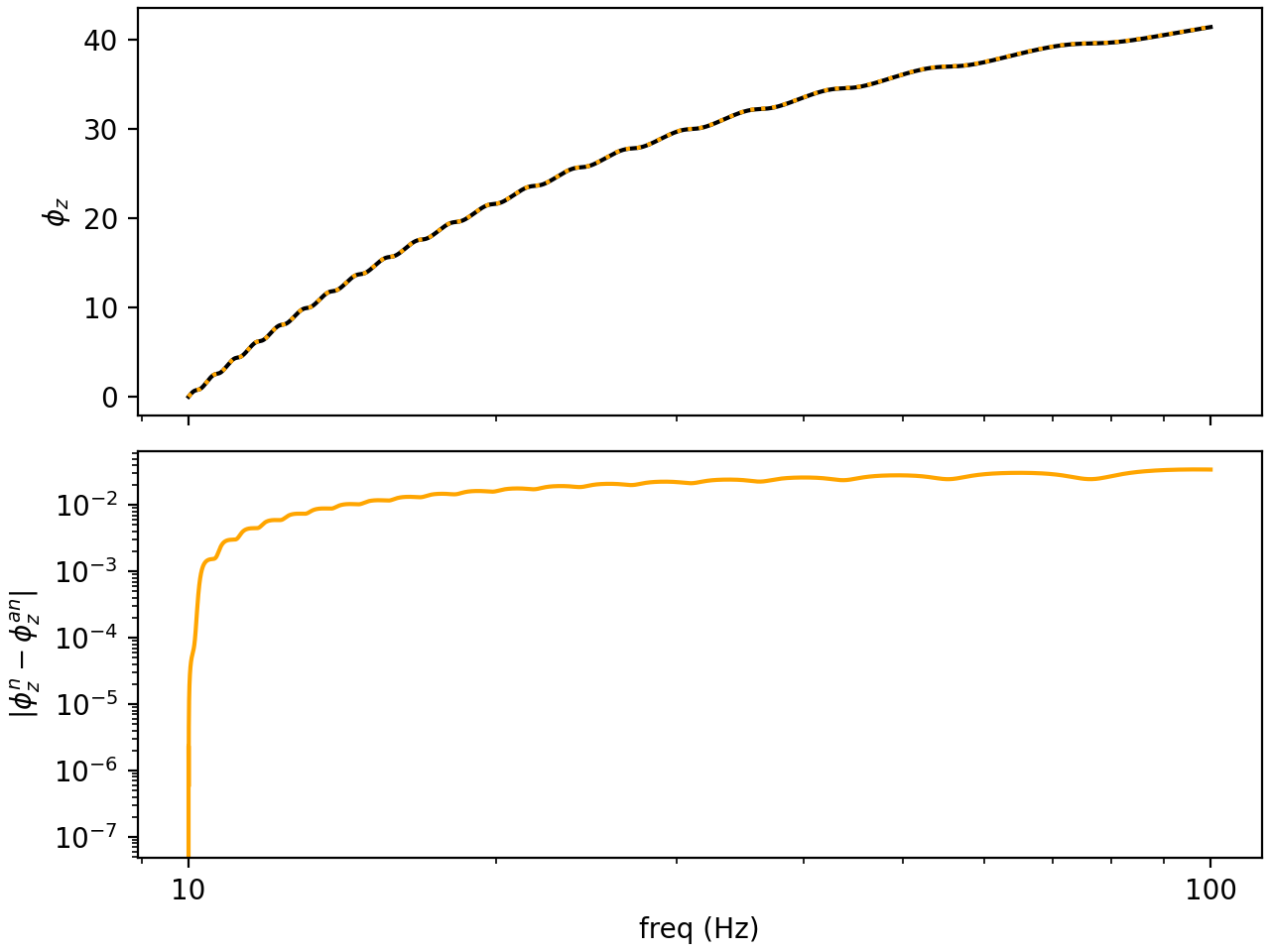}
		\caption[short]{NSBH Binary $\phi_z$}
	\end{subfigure}
	\caption[short]{Plots of the $\phi_z$ for the NSNS binary (a) and the NSBH binary (b), in black is the numerically evolved system, while orange is the system evolved with our method.  The top panel is the actual values of $\phi_z$ while the bottom panels are the absolute error between the value produced by the numerical evolution and that produced by our method.}
	\label{fig:numerical-vs-ourmethod-phiz}
\end{figure}

The absolute error for $\phi_{z}$ at low frequencies is $O(10^{-2})$ radians and at higher frequencies $ O(10^{-1})$ radians for the NSNS system, while for the NSBH system is maximally $O(10^{-2})$ radians, as shown in Fig.~\ref{fig:numerical-vs-ourmethod-phiz}. This error is dominated by its secular part, because of the complete lack of sharp dips (except at the lowest frequency, but this corresponds to the frequency at which we initialized the system).  The main source of this secular error is not obvious; there are two potential candidates. The first is the ``m=0'' approximation, by its very nature it produces errors of order $O(m)$ (which is equivalent to $O(y^2)$).  The second is the observed secular error in $\delta\chi$ and $\chi_{\text{eff}}$, which are also suppressed by $O(y^2)$. Since both error sources contribute at the same order, it is difficult to ascertain which one dominates this error budget.

\begin{figure}[H]
	\centering
	\begin{subfigure}{.5\textwidth}
		\centering
		\includegraphics[width=0.9\textwidth]{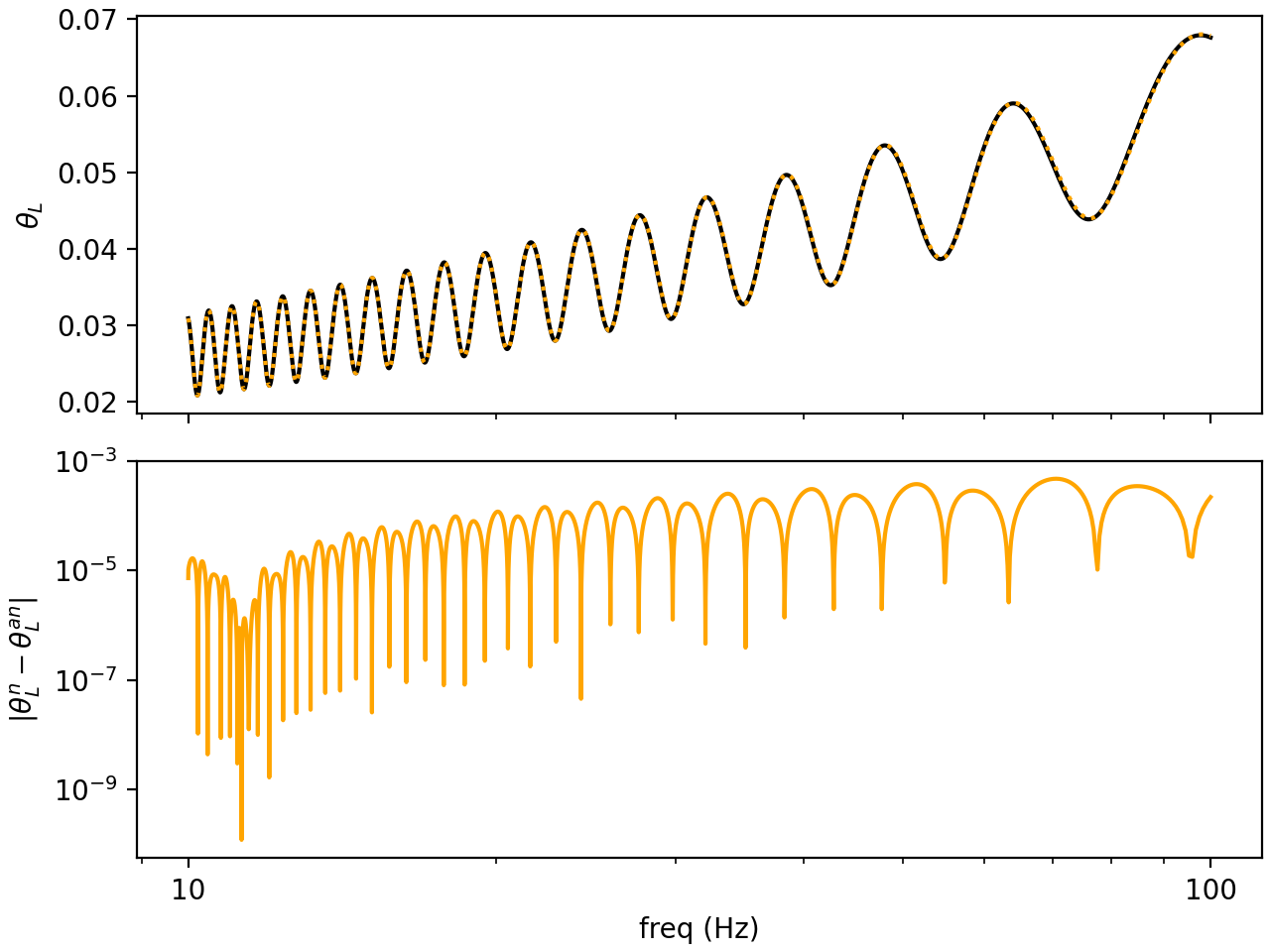}
		\caption[short]{NSNS Binary $\theta_L$}
	\end{subfigure}%
	\begin{subfigure}{.5\textwidth}
		\centering
		\includegraphics[width=0.9\textwidth]{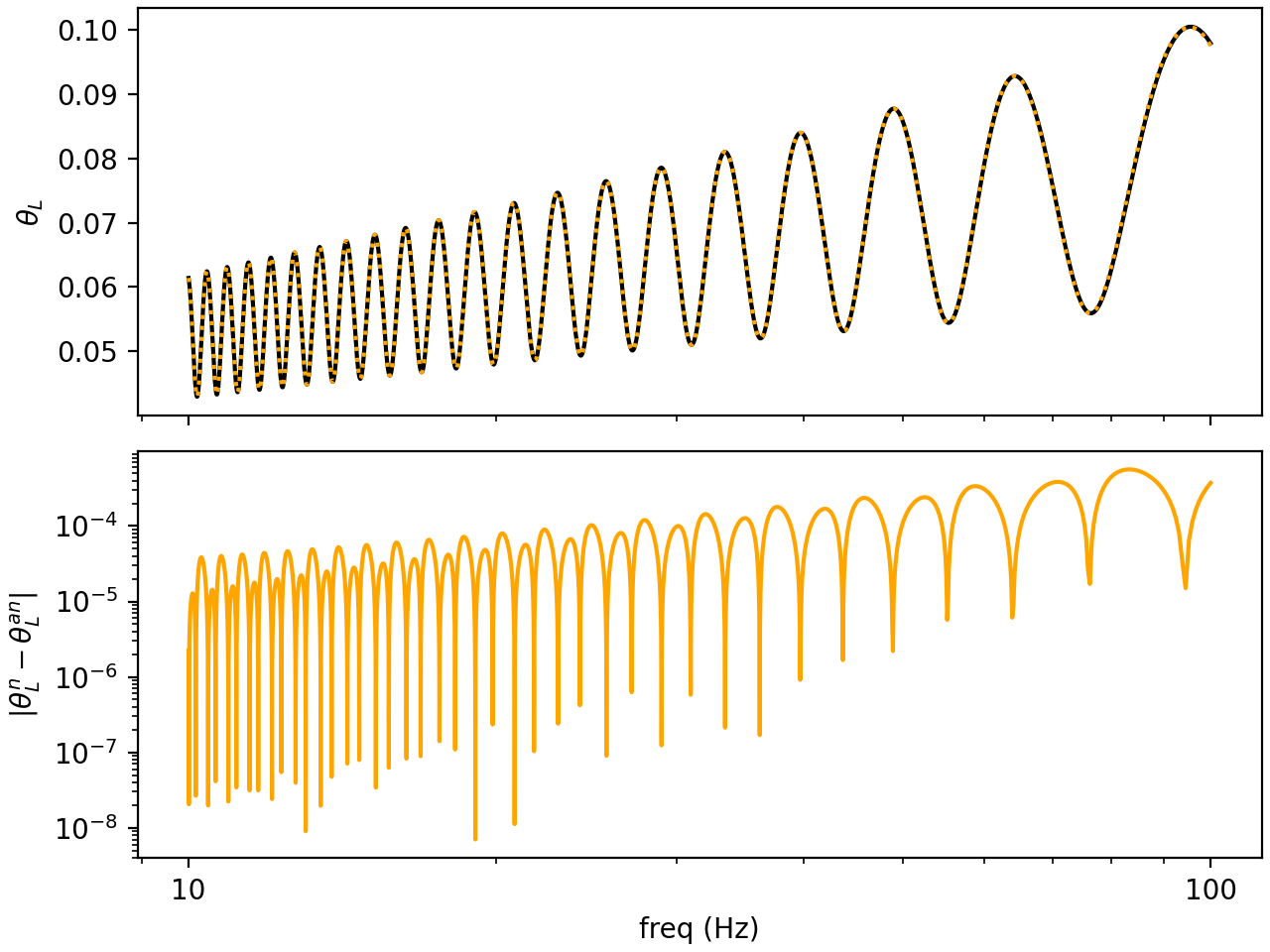}
		\caption[short]{NSBH Binary $\theta_L$}
	\end{subfigure}
	\caption[short]{Plots of $\theta_L$ as a function of frequency for the NSNS binary (a) and the NSBH binary (b). The black line shows the numerically evolved system, while orange shows the system evolved with our method. The top panel displays the actual values of $\theta_L$ while the bottom panels are the absolute error between the value produced by the numerical evolution and that produced by our method.}
	\label{fig:numerical-vs-ourmethod-thetaL}
\end{figure}

Finally, the error for $\theta_L$ is plotted in Fig.~\ref{fig:numerical-vs-ourmethod-thetaL}. This error is periodic and consequently one might be tempted to assume that its main source is the non-inclusion of higher terms in the Fourier series directly.  However, this is not the case, as the error has the same frequency as the precession, if they were from the non inclusion of these terms directly they would correspond to oscillations at double or triple the frequency. Hence, the major contributing source of error is small errors in the amplitude of $\delta\chi$.

To evaluate the accuracy of this waveform we evaluate the mismatch for two sets of systems:\\

The first set to evaluate the effectiveness for general NSNS systems, we evenly space the first mass between $[1.8,2.6]M_\odot$, and for the second we distribute its mass evenly between $1.0 M_\odot$ and $0.9 m_1$ (to ensure that the mass ratio is not equal to one).  We set the first spin to be $\chi_1 = 0.7$, and the second to be $\chi_2 = 0.2$, we then evenly distribute the angles $\theta_{s1}$,  $\theta_{s2} $, and $\phi_{s2}$ every $20^{\circ}$ over the sphere, evolving every system from 10-100Hz.\\

The second set of systems is more restricted in order to explore how much error is produced for models similar to GW190814, since this is the system of interest to us.  In particular, we restrict the masses to $m_1 = 23M_\odot$ and $m_2 = 2.6 M_\odot$, and the spins to $\chi_1 = \chi_2 = 0.6$.  Similarly we distribute the angles $\theta_{s1}$,  $\theta_{s2} $, and $\phi_{s2}$ every $20^{\circ}$ over the sphere, evolving every system from 10-100Hz.\\

\begin{figure}[H]
	\centering
	\begin{subfigure}{.5\textwidth}
		\centering
		\includegraphics[width=0.9\textwidth]{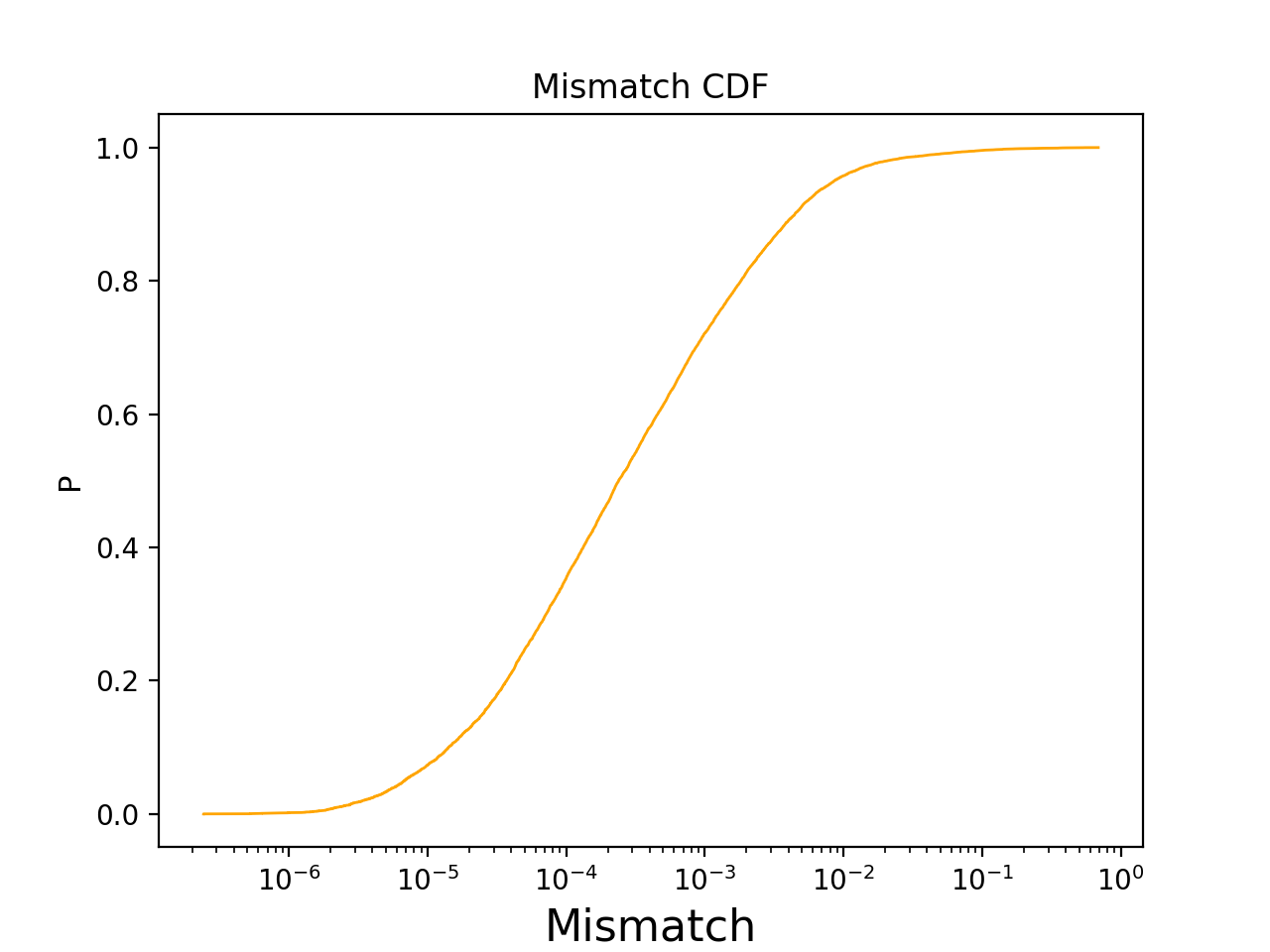}
		\caption[short]{NSNS Binary}
	\end{subfigure}%
	\begin{subfigure}{.5\textwidth}
		\centering
		\includegraphics[width=0.9\textwidth]{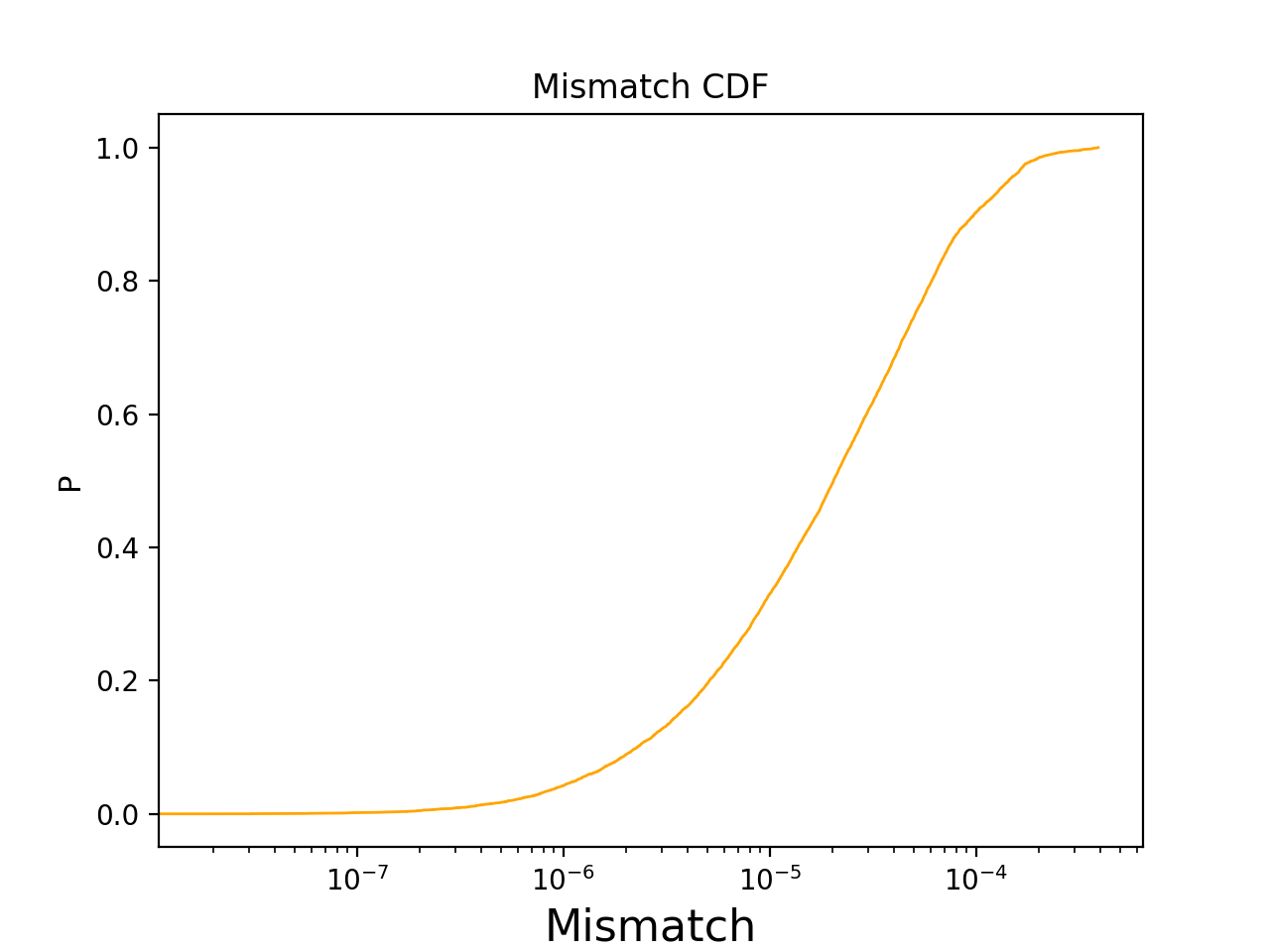}
		\caption[short]{NSBH Binary}
	\end{subfigure}
	\caption[short]{Plots of the CDF for mismatch between the numerically evolved NSNS/NSBH systems and the same systems evolved with our method.}
	\label{fig:numerical-vs-ourmethod-cdf}
\end{figure}

In the following we use the overlap, defined here as

	\begin{equation}\label{eq:S004-SS001-EQ001}
		\begin{split}
			{\rm overlap}(h_1,h_2) &= \frac{(h_1,h_2)}{\sqrt{(h_1,h_1)(h_2,h_2)}},\\
			 (h_1,h_2) &= 4 \rm Re \int_{f_{min}}^{f_{max}} h_1(f) h_2^*(f) df,
		\end{split}
	\end{equation}

where the waveform phase is aligned at the initial frequency.  This is done, since it will be a more conservative estimate of the accuracy than performing matched filterring.  The cumulative distribution functions (CDF) of the mismatch (one minus the overlap) between the waveform produced via our method and the waveform produced via the full numerical evolution are shown in Fig.~\ref{fig:numerical-vs-ourmethod-cdf}.  

For the set of binary neutron star systems we examine we find that only 1.35\%(7.44\%) have a fidelity worse than 0.965(0.994). This is the same order of magnitude as in \cite{Chatziioannou:2017tdw}. We have investigated several of the higher mismatch cases, and find that the main cause of these high mismatch cases was the non-inclusion of higher terms in the Fourier series of our ``m=0" approximation. This is because in these higher mismatch cases, $\theta_L$ drifts close to zero at its minimum.  A small error in the value of $\theta_L$ at this point then introduces larger effects in the secular evolution of $\phi_z$.  When the system precesses away from $\theta_L$ being close to zero, this then poses an issue and in turn contributes to the mismatch.  Adding more terms in the approximation would make this error smaller, in principal allowing us to reduce this mismatch.  In the NSBH case no systems we examined have overlap worse than 0.999\%.

\subsection{Differences between BNS/NSBH and BBH Waveforms}
\label{sec:effect-kappa}

Having established the excellent accuracy of our method with mismatch less than 3.5\% for 98.65\% in the more extreme case of NSNS systems (and an order of magnitude better accuracy in BHNS systems), here we study the qualitative effects of the spin-induced quadrupole moment on the evolution. In order to do so, we compare the two systems in Sec.~\ref{sec:effectiveness} to a binary with two black holes (BHBH) with the same parameters, except with $\kappa_1=1=\kappa_2$.
\\

Here we follow a similar procedure as in the previous section, using the same two sample configurations to evolve the NSNS/NSBH systems and then evolve the same system as though they were a BHBH system (with the other parameters all the same).\\

\begin{figure}[H]
	\centering
	\begin{subfigure}{.5\textwidth}
		\centering
		\includegraphics[width=0.9\textwidth]{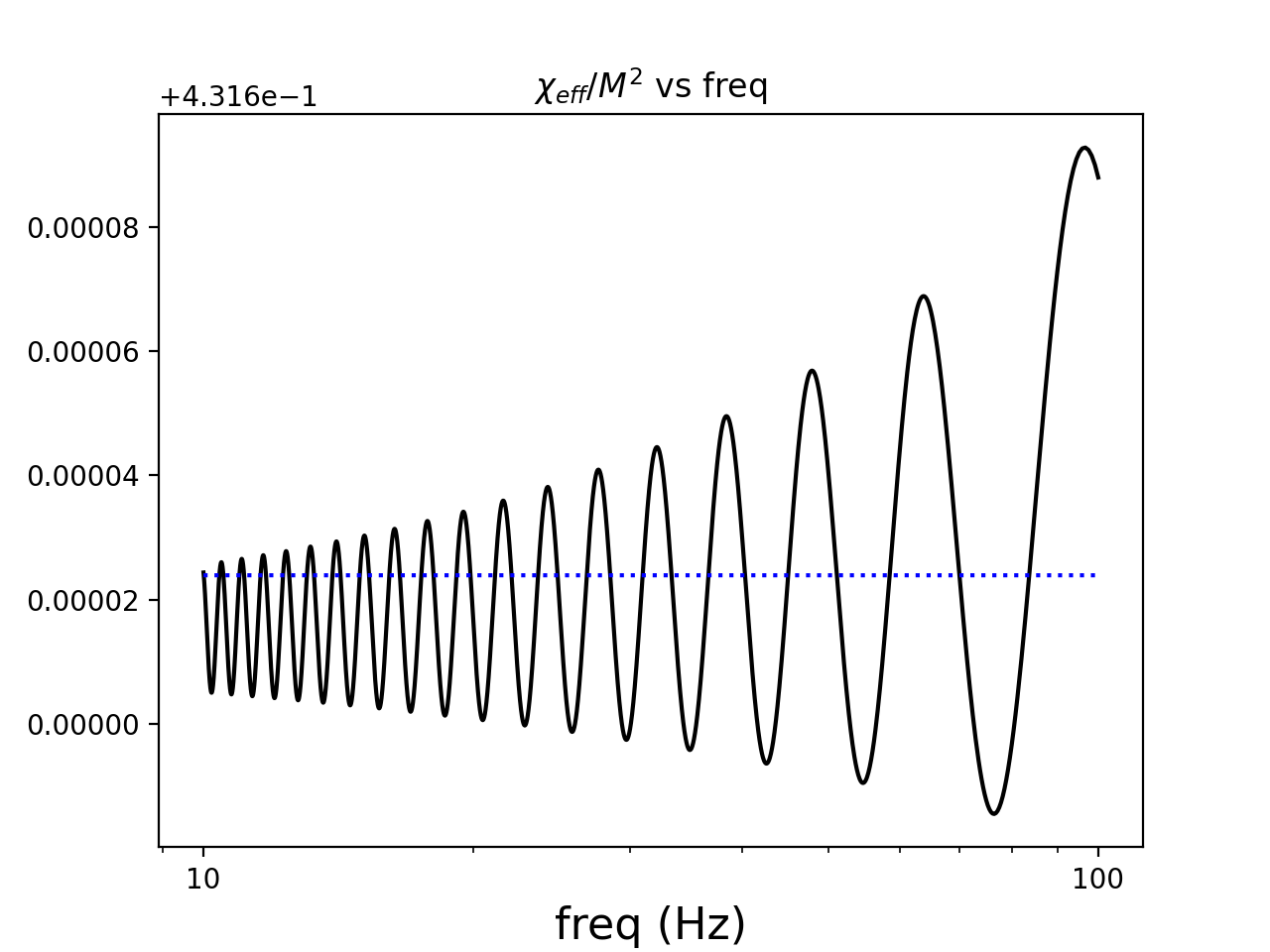}
		\caption[short]{NSNS binary}
	\end{subfigure}%
	\begin{subfigure}{.5\textwidth}
		\centering
		\includegraphics[width=0.9\textwidth]{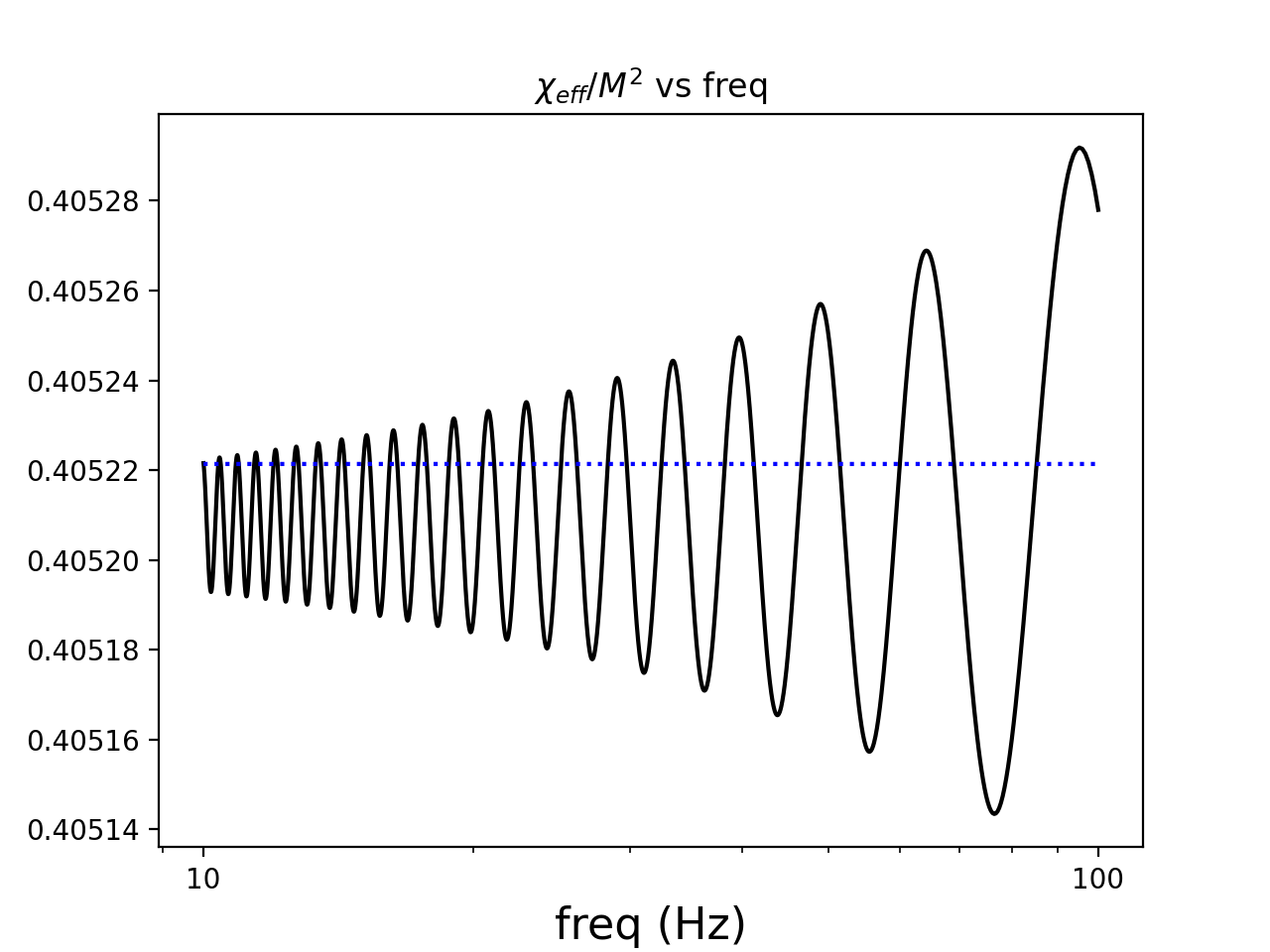}
		\caption[short]{NSBH binary}
	\end{subfigure}
	\begin{subfigure}{.5\textwidth}
		\centering
		\includegraphics[width=0.9\textwidth]{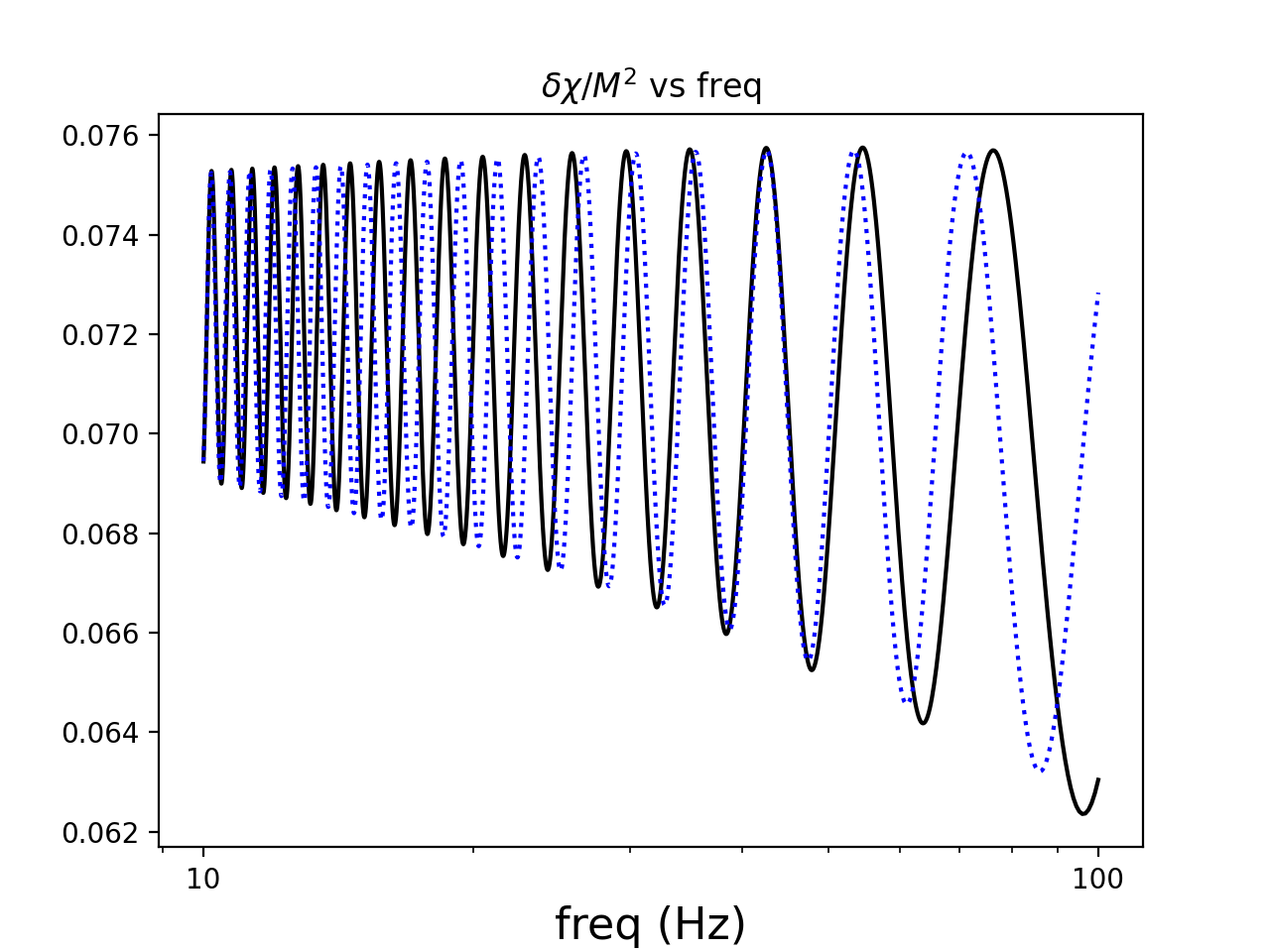}
		\caption[short]{NSNS binary}
	\end{subfigure}%
	\begin{subfigure}{.5\textwidth}
		\centering
		\includegraphics[width=0.9\textwidth]{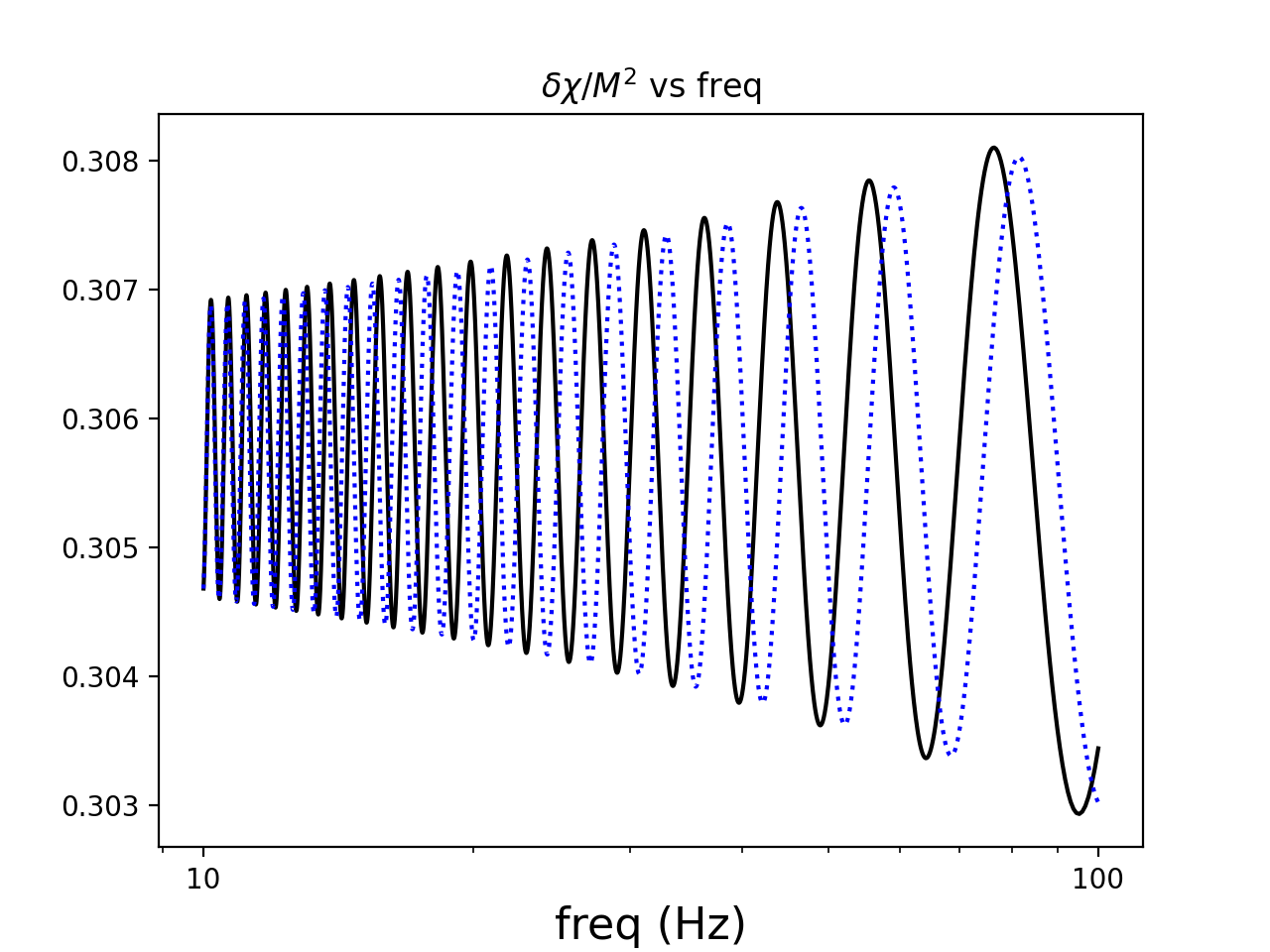}
		\caption[short]{NSBH binary}
	\end{subfigure}%
	\caption[short]{Plot of $\chi_{\text{eff}}/M^2$ vs frequency for the NSNS (NSBH) binary in panel a (b), and $\delta\chi/M^2$ for the NSNS (NSBH) binary in panel c (d).  In each plot the black curve is the numerically evolved neutron star system, while the blue curves are the corresponding BBH systems.}
	\label{fig:effect-kappa-chi}
\end{figure}

First, the effect of the quadrupole moment on $\chi_{\text{eff}}$ and $\delta\chi$ is shown in Fig.~\ref{fig:effect-kappa-chi}. Qualitatively the largest difference is the introduction of oscillations to $\chi_{\text{eff}}$, which are absent in the BHBH case. So one would expect the corresponding correction to be the same PN order as the amplitude of $\chi_{\text{eff}}$'s oscillations, which are $O(y^2)$. If the amplitude alone were modified this would introduce a relatively small periodic error. As we will see, the main contributor to the error is the modification to the precession frequency, observed via the difference in oscillation frequency between the blue and black curves.  Even small changes in frequency can introduce large differences in the final result.  Also note the absence of large secular differences in either $\chi_{\text{eff}}$ or $\delta\chi$, this will be important for later.

\begin{figure}[H]
	\centering
	\begin{subfigure}{.5\textwidth}
		\centering
		\includegraphics[width=0.9\textwidth]{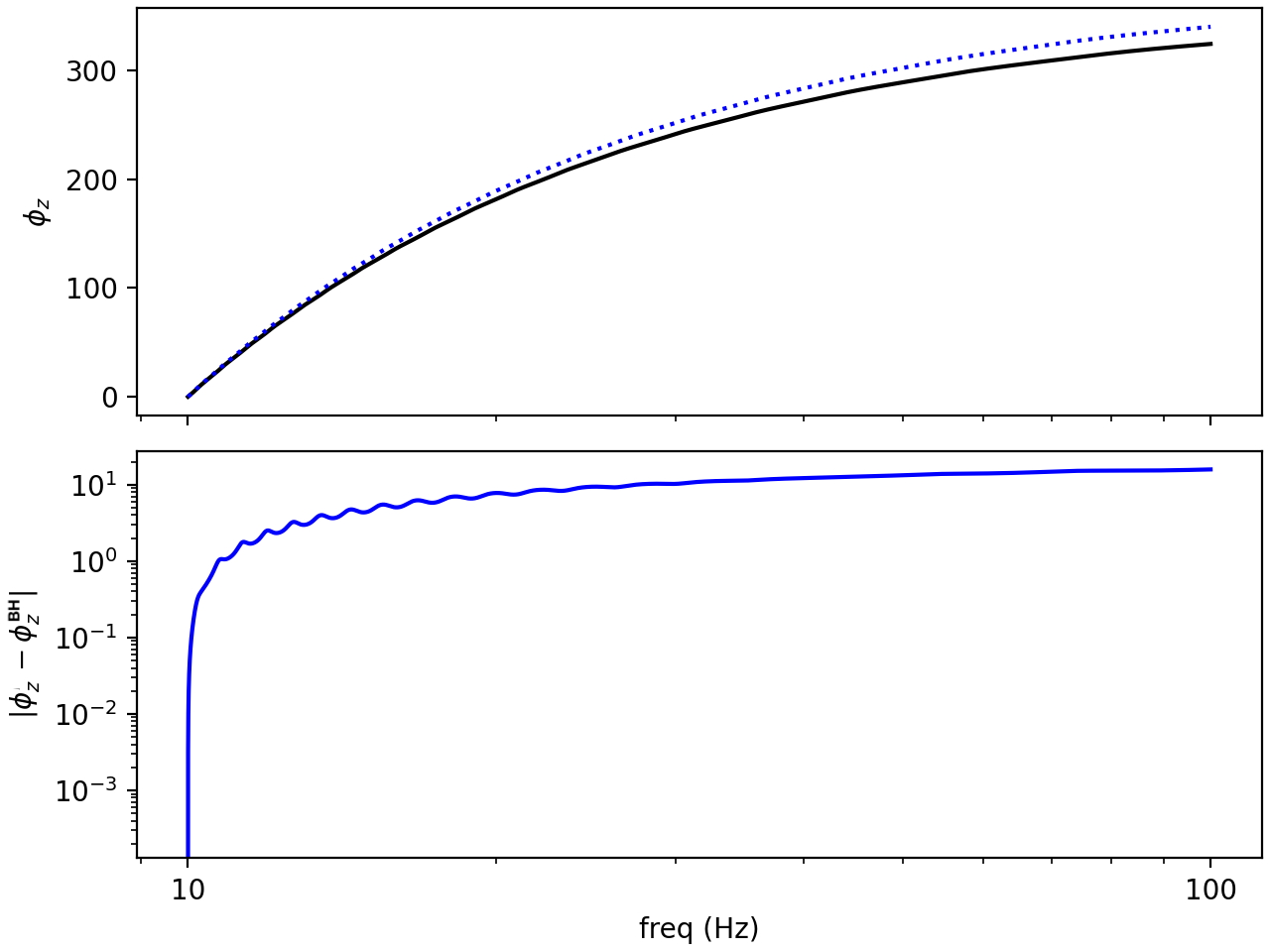}
		\caption[short]{NSNS Binary $\phi_z$}
	\end{subfigure}%
	\begin{subfigure}{.5\textwidth}
		\centering
		\includegraphics[width=0.9\textwidth]{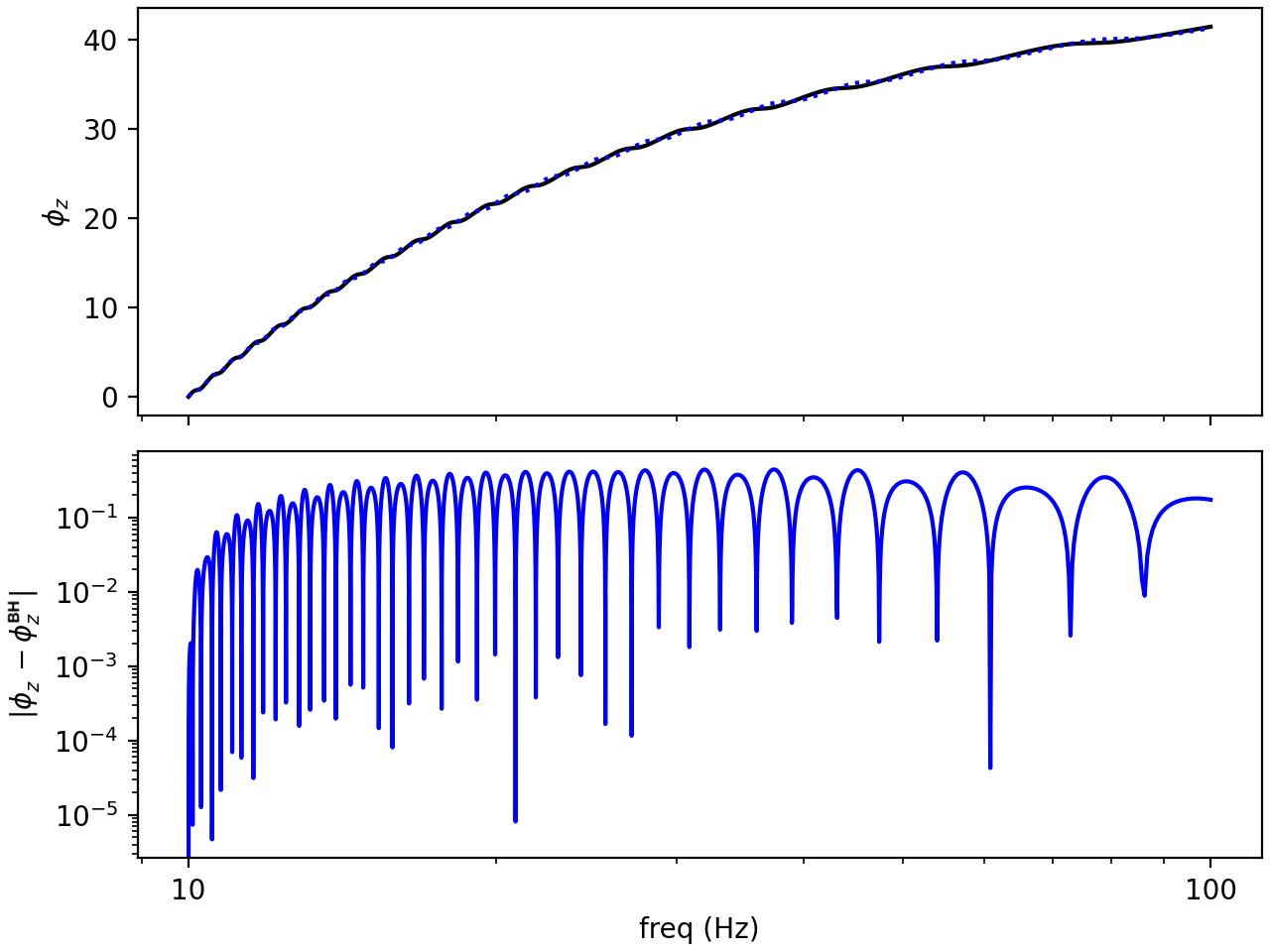}
		\caption[short]{NSBH Binary $\phi_z$}
	\end{subfigure}
	\caption[short]{Plots of the $\phi_z$ for the NSNS binary (left) and the NSBH binary (right), in black is the numerically evolved system, while blue is the equivalent BBH system.  The top panels show the actual values of $\phi_z$, while the bottom panels show the absolute difference between the NSNS/NSBH and BHBH binaries. }
	\label{fig:effect-kappa-phiz}
\end{figure}

Next, we turn to the angle $\phi_z$ in Fig.~\ref{fig:effect-kappa-phiz}. The difference between the NSNS and BHBH case is large enough that it is directly visible, and the difference accumulates to several radians. The source of this secular effect is the corrections to the terms in the derivative of the average, Eq.~\eqref{eq:eq:S003-SS006-EQ003}, involving $Q_{11}$-$Q_{16}$. Thus, the effects of the spin-induced quadrupole moment are important and introduce potentially huge errors to the waveform if left unaccounted for, even for smaller quadrupole moment constants between 1.1-1.5\\

The difference between the NSBH and BHBH systems is not secular but periodic in nature, characterized by the sharp, repeating dips.  This is because the secular part for a NSBH system is suppressed by the mass ratio.  Intuitively, this makes sense: the larger body's spin-induced quadrupole moment affects the waveform more. In the NSNS case, both masses are similar, whereas the masses can differ significantly in a NSBH binary. For the NSBH case we considered, the mass ratio is large so that the secular part of the evolution of $\phi_z$ is little affected by the quadrupole moment of the light neutron star.\\

\begin{figure}[H]
	\centering
	\begin{subfigure}{.5\textwidth}
		\centering
		\includegraphics[width=0.9\textwidth]{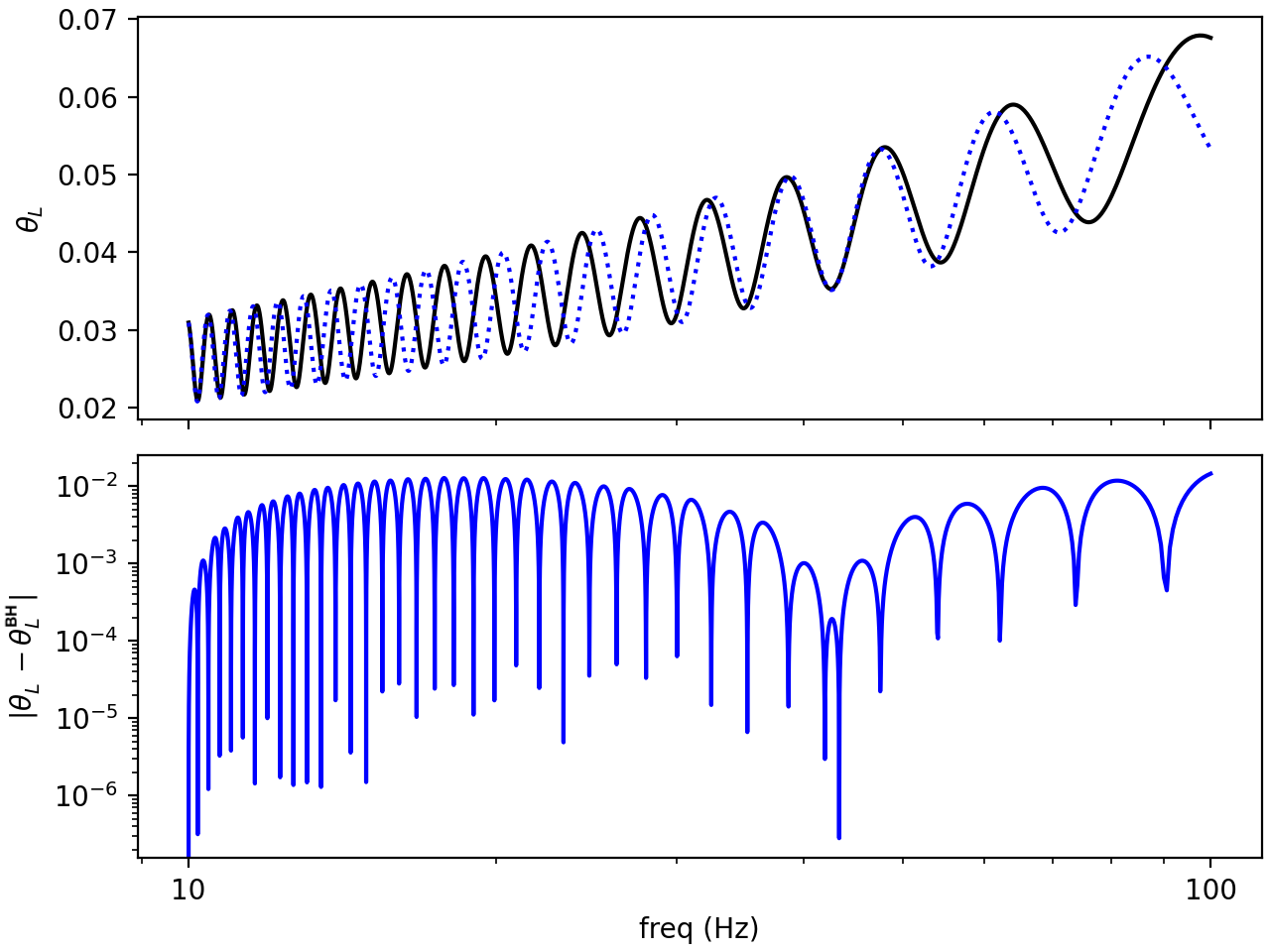}
		\caption[short]{NSNS Binary $\theta_L$}
	\end{subfigure}%
	\begin{subfigure}{.5\textwidth}
		\centering
		\includegraphics[width=0.9\textwidth]{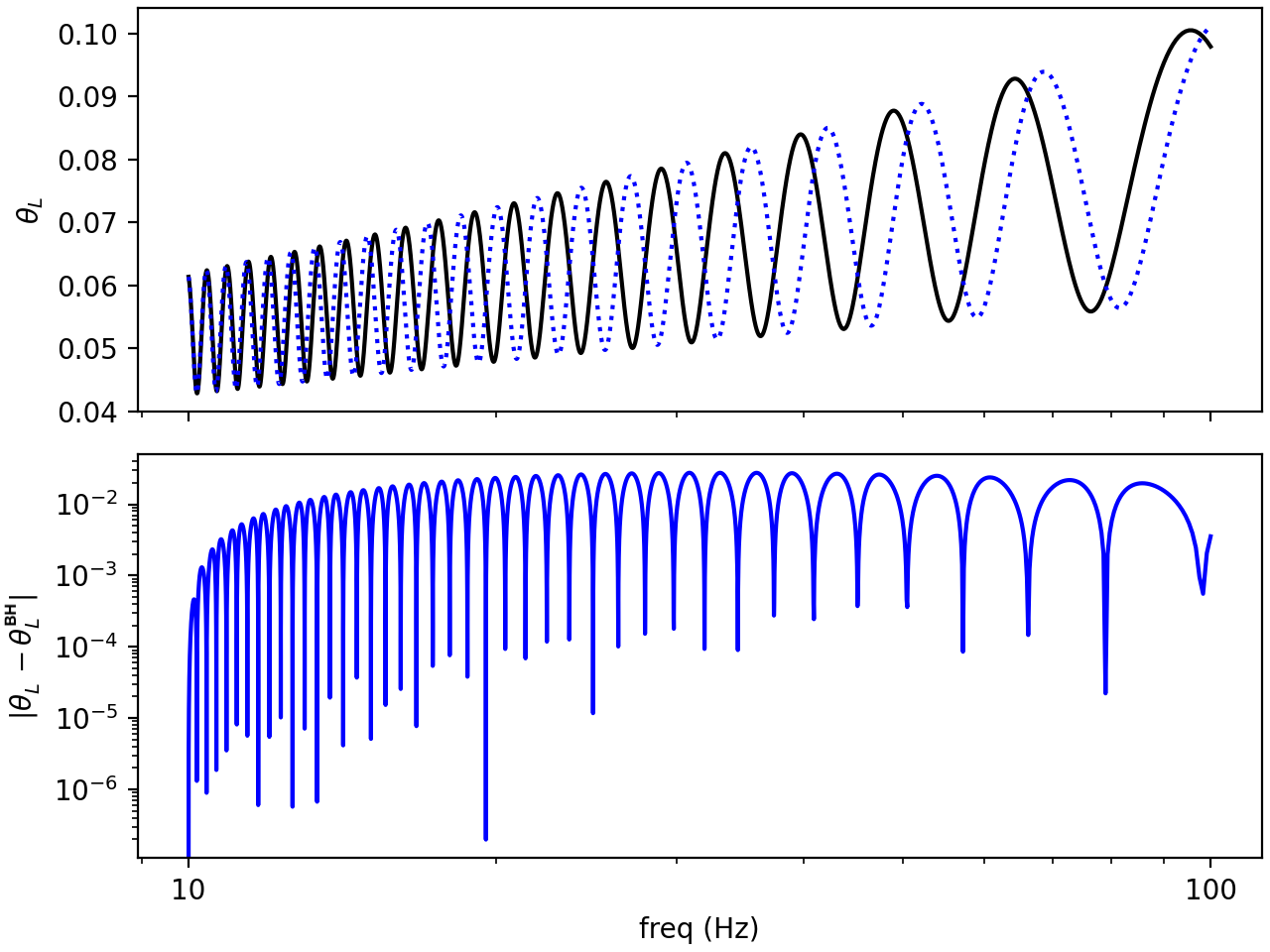}
		\caption[short]{NSBH Binary $\theta_L$}
	\end{subfigure}
	\caption[short]{Plots of the $\theta_L$ for the NSNS binary (a) and the NSBH binary (b), in black is the numerically evolved system, while blue is the equivalent BBH system.  The top panels are the actual values of $\theta_L$ while the bottom panels are the absolute difference between the value produced by the neutron star system and that of a BBH. }
	\label{fig:effect-kappa-thetaL}
\end{figure}

Finally, we investigate the influence of the quadrupole moment on $\theta_L$ in Fig.~\ref{fig:effect-kappa-thetaL}. The quadrupole moment leaves an imprint that is periodic. This is as expected, since we observe no significant secular imprint of $\kappa \neq 1$ on either $\chi_{\text{eff}}$ or $\delta\chi$.  The only other possible quantity that could introduce a secular effect to $\theta_L$ is $J$. However the secular effect in $J$ is largely determined by that of $\chi_{\text{eff}}$ and $\delta\chi$, so if the secular effect in $\chi_{\text{eff}}$ or $\delta\chi$ is already small, the secular effect in $J$ will also be small.  As a result, since the secular effect of the quadrupole moment is small in any of the quantities that determine $\theta_L$, there is little secular effect on $\theta_L$.\\

Since $\phi_z$ and $\theta_L$ play an important role in waveform generation, these results imply that precession is very important for parameter estimation.

To accurately assess the measurability of these effects we will use a more accurate measure of the mismatch that reflects what would be used in practice.  First, we define the fidelity:

	\begin{equation}\label{eq:S004-SS001-EQ001}
		\begin{split}
			{\rm F}(h_1,h_2) &= \max_{\phi} \frac{(h_1,h_2)}{\sqrt{(h_1,h_1)(h_2,h_2)}},\\
			 (h_1,h_2) &= 4 \rm Re \int_{f_{min}}^{f_{max}} \frac{h_1(f) h_2^*(f)}{S_n(f)} df,
		\end{split}
	\end{equation}

where now we perform matched filtering, maximizing the fidelity over the initial phase (or minimizing the mismatch), and we also include the advanced LIGO zero-detuned, high power noise sensitivity \cite{LIGOScientific:2014pky}.  So that we can accurately determine the magnitude of these effects with everything accounted for, we include the higher order terms in $dy/dt$ and the contributions of the quadrupole moment constant to the non-spinning part of the waveform \cite{Harry:2018hke}.  The cumulative distribution functions (CDF) of the mismatch (one minus the fidelity) between the BHNS/BNS waveform and the corresponding BBH waveform are shown in Fig.~{fig:effect-kappa-cdf}.  We fix the masses and quadrupole moment constants to those indicated in each panel in the figure, the dimensionless spins are fixed to $\chi_1=0.7$ and $\chi_2=0.6$ for all samples.  We take 1000 samples with these inputs with randomly distributed isotropic spin angles and inclination angle.

\begin{figure}[H]
	\centering
	\begin{subfigure}{.5\textwidth}
		\centering
		\includegraphics[width=0.9\textwidth]{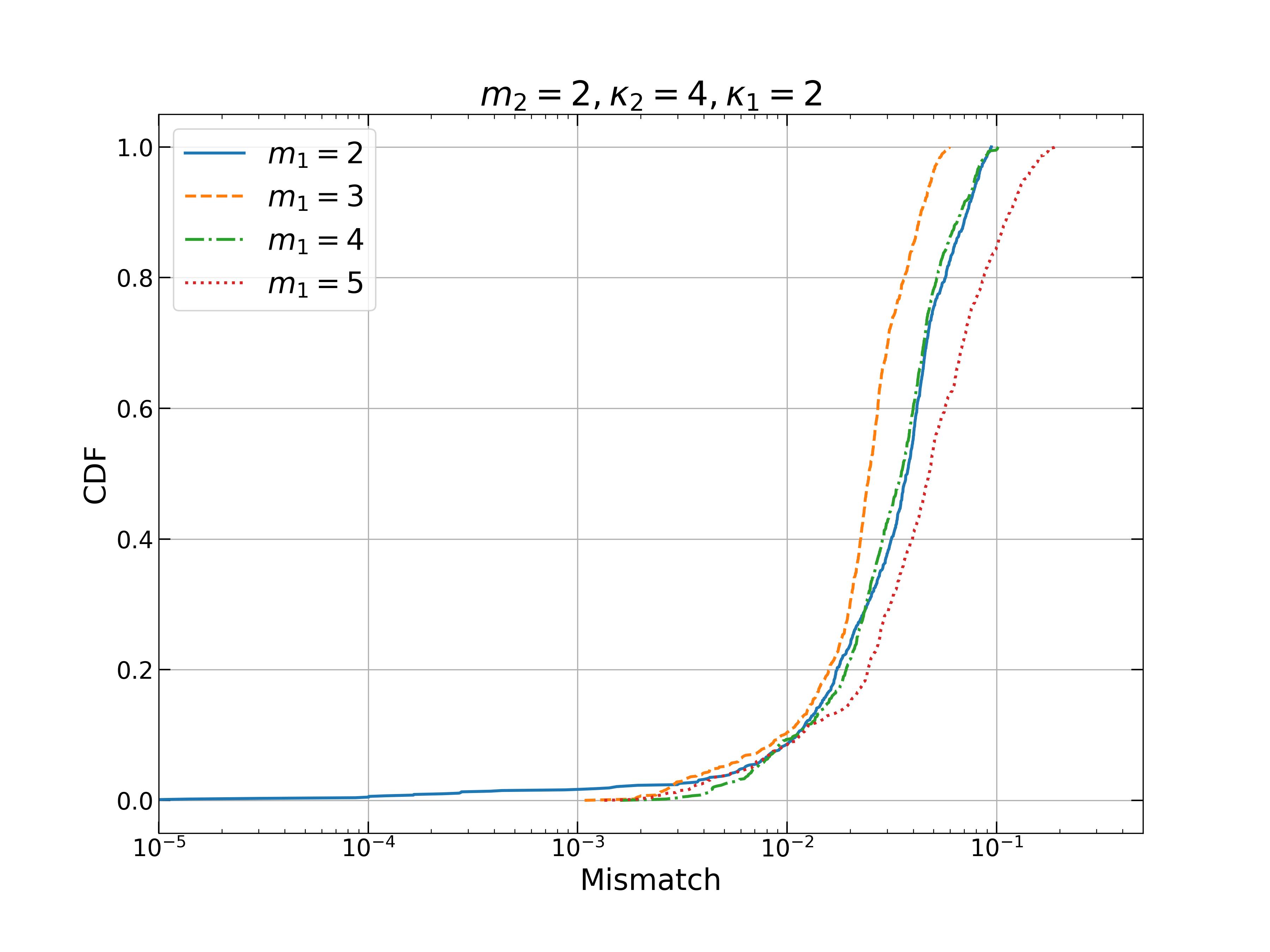}
		\caption[short]{NSNS Binary}
	\end{subfigure}%
	\begin{subfigure}{.5\textwidth}
		\centering
		\includegraphics[width=0.9\textwidth]{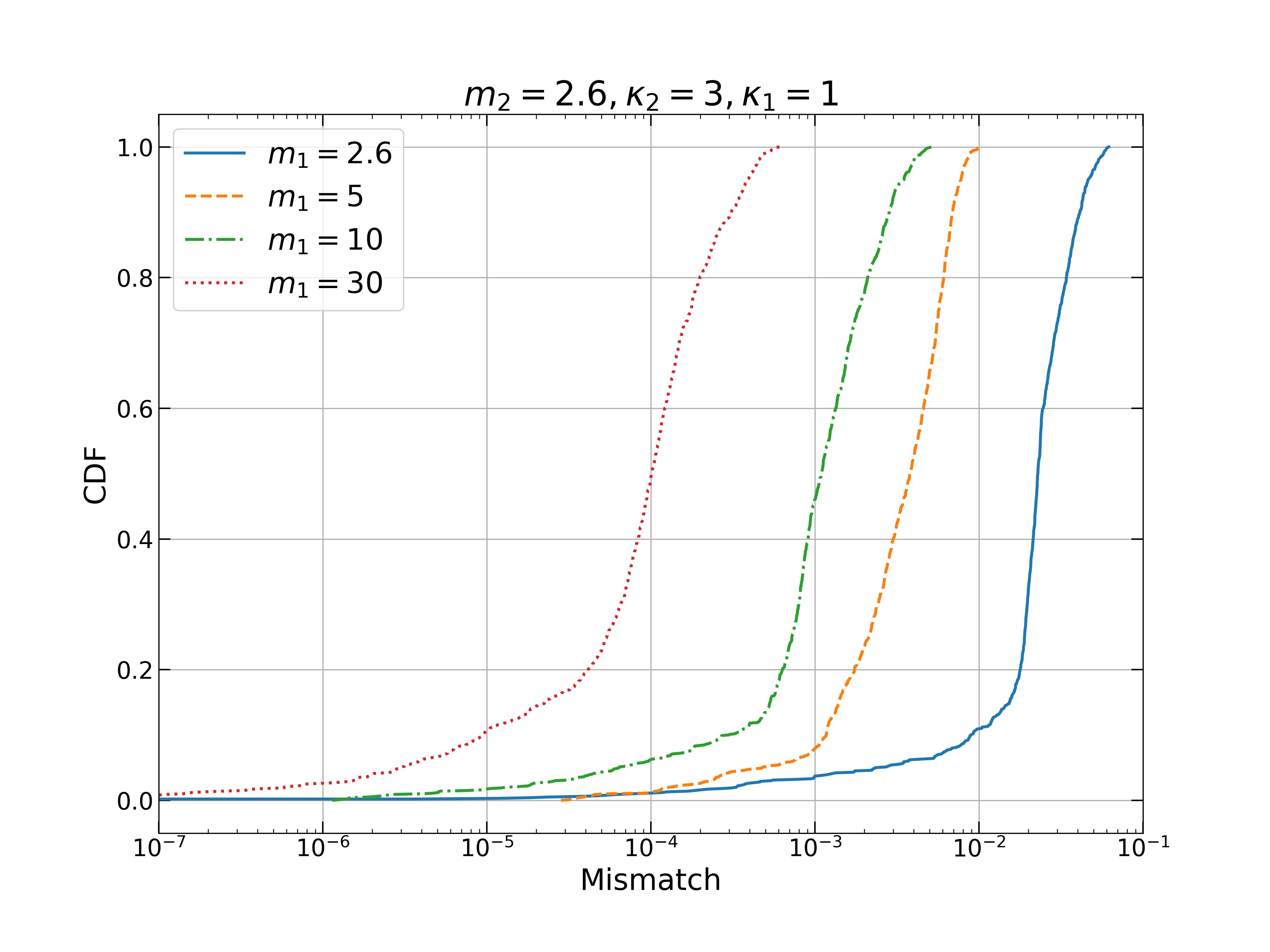}
		\caption[short]{NSBH Binary}
	\end{subfigure}
	\caption[short]{Plots of the CDF for mismatch between the numerically evolved NSNS/NSBH systems and the corresponding numerically evolved BHBH system.}
	\label{fig:effect-kappa-cdf}
\end{figure}

For the neutron star binary, this effect produces a large mismatch, but for the NSBH system this effect is more difficult to measure.  The explanation is that the secular effect is suppressed by the mass ratio, which is relatively small in the higher mass ratio BHNS cases we examine.  We can see this directly in the second panel of Fig.~{fig:effect-kappa-cdf}, where for gradually larger mass ratios the average mismatch between the BNS/NSBH system and the corresponding BBH system decreases.\\

By examining which cases have higher/lower mismatch, we find the quadrupole moment constant is more measurable for somewhat aligned/anti-aligned spins.  This can be understood intuitively from the precession equations (Eq.~\eqref{eq:eq:S00B-SS001-EQ008}), while the precession is larger for spins perpendicular to the orbital angular momentum, this is not the hypothetical best place to find the \emph{difference} in the precession.  This is because the correction terms scale with the projection of the spin onto the orbital angular momentum.  This means that the best systems to measure the difference between the black hole precession and neutron star precession are systems for which their spins are misaligned enough for the precession to be measurable, but somewhat aligned otherwise to maximize the difference between the neutron star and black hole precession.\\

 These results are consistent with those reported in studies of the effects on the waveform \cite{Krishnendu:2017shb,Krishnendu:2019tjp,Chia:2022rwc} in several qualitative ways.  Firstly, the spin precession effect is larger for more aligned spins, but as we have shown, if there is a slight misalignment in the spins the effects on the precession cannot be ignored if one wants accurate waveforms.  Secondly, the effects are more noticeable when the first spin is larger, because the secondary's spin affects the waveform less.  Finally, for the dimensionless spin magnitudes used here, this effect is in principle measurable.\\
 
 The tidal heating effect, which comes from the energy and angular momentum on the black hole horizon, contributes at $2.5$ PN order to the waveform if the black hole is spinning \cite{Tagoshi:1997jy}, and $4$PN if the black hole is non-spinning \cite{Poisson:1994yf}. Since neutron stars do not have tidal heating effects, in principle the tidal heating effect can also be used to probe the nature of mass-gap objects. However, the calculation in \cite{Alvi:2001mx} shows that the tidal heating effect from an equal mass black hole (with high spin $a=0.998$) binary contributes to $\sim 0.05$ gravitational wave cycle, which is approximately $0.3$ rad in phase modulation. For unequal mass ratio binaries, the flux absorbed by the less massive black hole $M_2$ is roughly $M_2/M_1$ times smaller than the flux absorbed by $M_1$, if both black holes are spinning, and $(M_2/M_1)^2$, if both are non-spinning. As a result, replacing $M_2$ by a neutron star in a GW 190814-like binary has negligible  impact on the gravitational wave phase if only the tidal heating effect is considered. 
 

\section{Conclusion}
\label{sec:discussion}
We have presented  a new method to construct frequency domain waveforms for circular compact object binaries that include neutron stars.	
Because of the analytical treatment of the spin evolution equations, the spin variables are evolved on the radiation reaction timescale, which is
convenient to transform to the frequency domain. The new waveform is able to achieve more than an order of magnitude speed-up compared to
the one with fully numerical evolution of spins on the precession timescale. 
For generic mass ratio and spin configurations, the mismatch between the new waveform
and the one with fully numerical evolution of spins is $\le 3.5\%$ for 98.65\% of configurations examined. We have also investigated the difference between BHBH, NSNS and NSBH waveforms due to different spin-induced quadrupole moment constant values, assuming the same component mass and spins and neglecting possible tidal effects. We find that the maximum difference occurs for system with somewhat aligned or anti-aligned spins. In such cases the mismatch between BHBH and BNS waveforms can be approximately $\sim 10\%$, which is promising for detection  with Advanced LIGO/Virgo \cite{LIGOScientific:2014pky,VIRGO:2014yos} and LIGO A+ \cite{KAGRA:2013rdx} . For other spin configurations, the difference between the waveforms are rather small, which may require third-generation gravitational wave detectors to measure the spin-induced quadrupole moment constant. For BNS systems, if the neutron stars have significant spins, the mismatch may reach $5\%$ for a good fraction of spin-configurations. This is particularly interesting if in the future we discover BNS systems with two mass-gap objects.\\

There are several avenues for further developments. First, improving this waveform's speed and accuracy via a semi-analytic evolution.  This can be done by expanding the averages and amplitudes of $\delta\chi$ and $\chi_{\text{eff}}$ in a series in terms of the PN parameter $y$:
\begin{equation}\label{eq:eq:S00B-SS001-EQ001}
		\begin{split}
			\left< \delta\chi \right> &= \left< \delta\chi \right>_0 + \left< \delta\chi \right>_1 y + \left< \delta\chi \right>_2 y^2, \\
			G_{\delta\chi} &= G_{\delta\chi,0} + G_{\delta\chi,1} y + G_{\delta\chi,2} y^2, \\
			\left< \chi_{\text{eff}} \right> &=  \left< \chi_{\text{eff}} \right>_0 + \left< \chi_{\text{eff}} \right>_1 y + \left< \chi_{\text{eff}} \right>_2 y^2, \\
			G_{\chi_{\text{eff}} } &=G_{\chi_{\text{eff}} , 0 } + G_{\chi_{\text{eff}} , 1 } y + G_{\chi_{\text{eff}} , 2} y^2.
		\end{split}
	\end{equation}
This approach has the advantage that while there is a higher up front cost to initialize this system (requiring solving a system of linear equations), it makes the cost of calculating the amplitudes and averages at each step significantly lower and removes the need for root-finding altogether. The system of equations for these coefficients would be formed from the initial values of the amplitudes and averages, which can be found using a similar set of derivatives of $\delta\chi$ and $\chi_{\text{eff}}$ used to calculate the initial averages here.  The rest of the equations would be formed from derivatives of the roots of $\delta\chi$ and $\chi_{\text{eff}}$; we have discussed in an earlier section how to find such derivatives via the implicit function theorem.  To keep high accuracy in the waveform, $\psi$ and $\left< \phi_z\right>$ could potentially still be numerically evolved if necessary.\\

Normally speed versus accuracy is a balancing act: improving one typically harms the other. The approach taken in this paper provides the potential for improving both simultaneously.  The speed improvement we mentioned already and the accuracy can be further improved by including the next terms in the Fourier series in the ``m=0" approximation. The higher order terms improve the accuracy in two ways. First, including them make the estimation of the average and amplitude of $\delta\chi$ and $\chi_{\text{eff}}$ more accurate, thereby lowering the periodic effects in the error plots of $\theta_L$ which correspond to $1\times$ the precession frequency.  On top of this, the inclusion of the higher order terms should remove some of the error with double this frequency that is buried beneath this error. This improvement in $\theta_L$ will also improve the accuracy of $\phi_z$'s evolution, since the main source of error in $\phi_z$ is due to the error in the minimum of $\theta_L$.
\\

Currently, there is no waveform model that can simultaneously handle the evolution of eccentric orbits and precession. For neutron stars binaries this task is particularly complicated as tidal excitations may include multiple harmonics with eccentric orbits \cite{Yang:2018bzx,Yang:2019kmf}. However, this is (astro)physically important as some of the dynamically formed binaries may carry nonzero eccentricity and non-negligible precession at the same time. It is a promising direction to consider whether eccentric orbit evolution can be incorporated into the scheme discussed here.\\ 

Finally, it will be interesting to perform Bayesian parameter estimation with this waveform, either with the real data (such as GW 190814) or with artificial data (detector noise plus injected signal). Given that this is probably the best method to probe the nature of the mass-gap objects similar to the one found in GW 190814 ----if no electromagnetic counterparts are present --- it is important to assess the ability of using these precession waveforms to measure the spin-induced quadrupole moment constant in compact binaries with respect to various detector sensitivities. We are currently performing such analysis, the results will be reported in a future publication. \\

\vspace{0.2cm}
{\bf Acknowledgement.}
We thank Mohammed Khalili for reading over the manuscript and providing many helpful comments. We thank Reed Essick and the Perimeter LSC discussion group for useful  discussions. M. L., Z. L. and H. Y. are supported by the Natural Sciences and
Engineering Research Council of Canada and in part by
Perimeter Institute for Theoretical Physics. Research at
Perimeter Institute is supported in part by the Government
of Canada through the Department of Innovation, Science
and Economic Development Canada and by the Province of
Ontario through the Ministry of Colleges and Universities.


	\newpage
	\appendix
	
	\section{Coefficients of $d\delta\chi/dt$ with behaviour separated}
	\label{app:coefficients}
	
	The coefficients in Eq.~\eqref{eq:S002-SS001-EQ001}-\eqref{eq:S002-SS001-EQ002} can be found in Klein \cite{Klein:2021jtd}, but are repeated here:
	\begin{subequations}
		\begin{align}	
			B &= \frac{y}{2\eta^2} \left[ -2\eta (J^2 - L^2 -L\chi_{\text{eff}} ) + \delta\mu (S_1^2 -S_2^2) - \delta\mu^2 (2L^2 + S_1^2 + S_2^2) \right],\\
			C &= \frac{y}{2\eta^2} \left\{ (1 + \delta\mu^2)\chi_{\text{eff}}(S_1^2 - S_2^2) + 2\delta\mu \left[ 2L(J^2 - L^2 - L\chi_{\text{eff}} ) - (2L+\chi_{\text{eff}})(S_1^2 + S_2^2 ) - \eta L \chi_{\text{eff}}^2 \right] \right\}\\
			D &=\frac{y}{2\eta^2} \left\{ -2( J^2 - L^2 - L\chi_{\text{eff}} )\left[ J^2 - L^2 -L\chi_{\text{eff}} - 2(S_1^2 +S_2^2) - \eta\chi_{\text{eff}}^2 \right] \right.\\
			&+\left. (S_1^2 - S_2^2) \left[ \delta\mu \chi_{\text{eff}}^2 - 2(S_1^2 - S_2^2) \right] - \chi_{\text{eff}}^2 (S_1^2 + S_2^2) \right\} \notag
		\end{align}
	\end{subequations}
	From these it follows that the coefficients with the behavior of $\chi_{\text{eff}}$ separated (see Eqs.~\eqref{eq:S002-SS001-EQ005}) are	
	\begin{subequations}
		\begin{align}	
			B_0 &= \frac{y}{2\eta^2} \left[ -2\eta (J^2 - L^2 ) + \delta\mu (S_1^2 -S_2^2) - \delta\mu^2 (2L^2 + S_1^2 + S_2^2) \right]\\
			B_1 &= 1\\~\nonumber \\ \StepSubequations
			C_0 &= \frac{2 \delta\mu}{\eta} \left[ J^2 - L^2 - S_1^2 - S_2^2 \right]\\
			C_1 &= \frac{y}{2\eta^2} \left[ (1 + \delta\mu^2) (S_1^2 - S_2^2) - 2\delta\mu (2L^2 + S_1^2 + S_2^2) \right]\\
			C_2 &= -\delta\mu\\~\nonumber \\ \StepSubequations
			D_0 &= \frac{-y}{\eta^2} \left( J^2 - L^2 - S_1^2 - 2S_1 S_2 - S_2^2 \right) \left( J^2 - L^2 - S_1^2 + 2S_1 S_2 - S_2^2 \right)\\
			D_1 &= \frac{2}{\eta} \left( J^2 - L^2 - S_1^2 - S_2^2 \right)\\
			D_2 &= \frac{y}{2\eta^2} \left[ 2\eta (J^2-L^2) - (2L^2 +S_1^2 + S_2^2) + \delta\mu (S_1^2 - S_2^2) \right]\\
			D_3 &= -1
		\end{align}
	\end{subequations}
	
	
	\section{Derivative of $\phi_z$ definitions}
	\label{app:phiz}
	
	We restate the derivative of $\phi_z$ here:
	
	\begin{equation}\label{eq:eq:S00A-SS002-EQ001}
		\begin{split}
			\frac{d\phi_z}{dt}  = \frac{1}{\sin^2(\theta_L)} \left[ \frac{d \hat{L}}{dt} \cdot \left( \hat{J} \times \hat{L} \right) \right].
		\end{split}
	\end{equation}
	
	By substituting the derivative $d\hat{L}/dt$, we can separate the behaviour into the terms that come from the quadrupole moment constant being $\neq1$ and those that correspond to the black hole portion, doing so gives:
	
	\begin{equation}\label{eq:eq:S00A-SS002-EQ002}
		\begin{split}
			\frac{d\phi_z}{dt}  =& -\frac{y^6}{2\sin^2(\theta_L)}\left[ \hat{L} \times (\mu_1 \vec{s}_1 + \mu_2 \vec{s}_2) \right] \cdot ( \hat{J} \times \hat{L} )\\
			&-\frac{3y^6}{2\sin^2(\theta_L)} \left[ 1 - y \chi_{\text{eff}} \right] \left\{  \hat{L} \times \vec{s}_1  +  \hat{L} \times \vec{s}_2  \right\} \cdot(  \hat{J} \times \hat{L} )\\
			&+ \frac{3y^7}{2\sin^2(\theta_L)} \left\{  (\kappa_1 - 1 )( \hat{L} \cdot \vec{s}_1 ) ( \hat{L} \times \vec{s}_1 ) + (\kappa_2 - 1 )( \hat{L} \cdot \vec{s}_2 )( \hat{L} \times \vec{s}_2 ) \right\} \cdot(  \hat{J} \times \hat{L} ).
		\end{split}
	\end{equation}
	
	By relying on the definition of $\vec{J}$ given in Eq.~\eqref{eq:eq:S00B-SS001-EQ007}, in addition to the definition of $\theta_L$ as the angle between $\vec{L}$ and $\vec{J}$ the first term can be simplified substantially
	
	\begin{equation}\label{eq:eq:S00A-SS002-EQ002}
		\begin{split}
			\frac{d\phi_z}{dt}  =& \frac{Jy^6}{2} - \frac{3y^6}{2\sin^2(\theta_L)} \left[ 1 - y \chi_{\text{eff}} \right] \left\{  \hat{L} \times \vec{s}_1  +  \hat{L} \times \vec{s}_2  \right\} \cdot(  \hat{J} \times \hat{L} )\\
			&+ \frac{3y^7}{2\sin^2(\theta_L)} \left\{  (\kappa_1 - 1 )( \hat{L} \cdot \vec{s}_1 ) ( \hat{L} \times \vec{s}_1 ) + (\kappa_2 - 1 )( \hat{L} \cdot \vec{s}_2 )( \hat{L} \times \vec{s}_2 ) \right\} \cdot(  \hat{J} \times \hat{L} ).
		\end{split}
	\end{equation}
	
	The latter terms correspond to neutron stars, while the former terms (alone) compose the entirety of the black hole case.  To simplify the derivative of $\phi_z$ we start by looking at its expression in the black hole case.  Substituting Eq.~\eqref{eq:S003-SS001-EQ001} gives	
	\begin{equation}\label{eq:eq:S00A-SS001-EQ001}
		\begin{split}
			\left[ \frac{d\phi_z}{dt} \right]^{\text{BH}} \approx& \frac{J y^6}{2} \left\{ Q_1+Q_2\sin(\psi) + \frac{(Q_3+Q_4\sin(\psi)) (Q_5 + Q_6\sin(\psi) + Q_7 \sin^2(\psi))}{ Q_8 Q_9( 1 - \frac{Q_{10}}{Q_8}\sin(\psi) )( 1 + \frac{Q_{10}}{Q_9}\sin(\psi) ) } \right\},
		\end{split}
	\end{equation}
	where
	\begin{subequations}
		\begin{align}	
			Q_1 &=1 + \frac{3}{2\eta}\left( 1 - y\left<\chi_{\text{eff}}\right> \right) \label{eq:eq:S00A-SS001-EQ002} \\
			Q_2 &= - \frac{3}{2\eta}y G_{\chi_{\text{eff}}} 	\label{eq:eq:S00A-SS001-EQ003} \\
			Q_3 &= - \frac{3(1+q)}{2q }\left( 1 - y\left<\chi_{\text{eff}}\right> \right) = \frac{1 - Q_1}{1 + q}\label{eq:eq:S00A-SS001-EQ004} \\
			Q_4 &= \frac{3(1+q)}{2q } y G_{\chi_{\text{eff}}} = \frac{-Q_2}{1+q} \label{eq:eq:S00A-SS001-EQ005} \\
			Q_5 &= 4(1-q) (S_1^2 - S_2^2) - (1+q)(\delta\mu \left<\delta\chi \right> + \left< \chi_{\text{eff}} \right> )(\delta\mu \left<\delta\chi \right> + (1-4\eta)\left< \chi_{\text{eff}} \right>) \label{eq:eq:S00A-SS001-EQ006} \\
			Q_6 &= - (1+q)\left[ (\delta\mu \left<\delta\chi \right> + \left< \chi_{\text{eff}} \right> )(\delta\mu G_{\delta\chi} + (1-4\eta)G_{\chi_{\text{eff}}}) + (\delta\mu G_{\delta\chi} + G_{\chi_{\text{eff}}} )(\delta\mu \left<\delta\chi \right> + (1-4\eta)\left< \chi_{\text{eff}} \right>) \right] \label{eq:eq:S00A-SS001-EQ007} \\
			Q_7 &= - (1+q)(\delta\mu G_{\delta\chi} + G_{\chi_{\text{eff}}} )(\delta\mu G_{\delta\chi} + (1-4\eta)G_{\chi_{\text{eff}}}) \label{eq:eq:S00A-SS001-EQ008} \\
			Q_8 &= 2\left<J\right> - \delta\mu \left<\delta\chi\right> - \left<\chi_{\text{eff}}\right> - 2L  \label{eq:eq:S00A-SS001-EQ009} \\
			Q_{9} &= 2\left<J\right> + \delta\mu \left<\delta\chi\right> + \left<\chi_{\text{eff}}\right> + 2L = 4\left<J\right> - Q_8 \label{eq:eq:S00A-SS001-EQ0010} \\
			Q_{10} &=   \delta\mu G_{\delta\chi} + G_{\chi_{\text{eff}}}. \label{eq:eq:S00A-SS001-EQ011}
		\end{align}
	\end{subequations}
	
	The rest of the terms, corresponding to the spin induced quadrupole moment of the neutron star are given by
	\begin{equation}\label{eq:eq:S00A-SS001-EQ012}
		\begin{split}
			\frac{d\phi_z}{dt} =  \left[ \frac{d\phi_z}{dt} \right]^{\text{BH}} + \frac{J y^6}{2} \left\{ \frac{  Q_{11} + Q_{12}\sin(\psi)  + ( Q_{13} + Q_{14}\sin(\psi) )( Q_{15} + Q_{16}\sin(\psi) + Q_{17}\sin^2(\psi) ) }{ Q_8 Q_9( 1 - \frac{Q_{10}}{Q_8}\sin(\psi) )( 1 + \frac{Q_{10}}{Q_9}\sin(\psi) ) } \right\},
		\end{split}
	\end{equation}
	where
	\begin{subequations}
		\begin{align}
			Q_{11} &= -3y \frac{J^2-L^2}{L\eta}\left[ (\kappa_2 - 1 )\mu_1( \left<\chi_{\text{eff}}\right> - \left<\delta\chi\right> )  + (\kappa_1 - 1 )\mu_2 ( \left<\chi_{\text{eff}}\right> + \left<\delta\chi\right> )  \right] \notag \\
			&+ 3y \frac{S_1^2-S_2^2}{L\eta}\left[ (\kappa_2 - 1 )\mu_1( \left<\chi_{\text{eff}}\right> - \left<\delta\chi\right> )  - (\kappa_1 - 1 )\mu_2 ( \left<\chi_{\text{eff}}\right> + \left<\delta\chi\right> )  \right]\label{eq:eq:S00A-SS001-EQ0013}\\
			Q_{12} &= -3y \frac{J^2-L^2}{L\eta}\left[ (\kappa_2 - 1 )\mu_1( G_{\chi_{\text{eff}}} - G_{\delta\chi} )  + (\kappa_1 - 1 )\mu_2 ( G_{\chi_{\text{eff}}} + G_{\delta\chi} )  \right] \notag \\
			&+ 3y \frac{S_1^2-S_2^2}{L\eta}\left[ (\kappa_2 - 1 )\mu_1( G_{\chi_{\text{eff}}} - G_{\delta\chi} )  - (\kappa_1 - 1 )\mu_2 ( G_{\chi_{\text{eff}}} + G_{\delta\chi} )  \right] \label{eq:eq:S00A-SS001-EQ0014} \\
			Q_{13} &= \frac{3}{2}(\delta\mu \left<\delta\chi \right> + \left< \chi_{\text{eff}} \right> ) \label{eq:eq:S00A-SS001-EQ0015} \\
			Q_{14} &= \frac{3}{2} (\delta\mu G_{\delta\chi} + G_{\chi_{\text{eff}}} ) \label{eq:eq:S00A-SS001-EQ0016} \\
			Q_{15} &= 4y A_{ \chi_{\text{eff}} , \delta\chi} \left<\delta\chi \right>^2 + 2( 4yA_{ \chi_{\text{eff}} , \chi_{\text{eff}} }\left< \chi_{\text{eff}} \right> + \mu_2 \kappa_1 - \mu_1 \kappa_2 + \delta\mu )\left<\delta\chi \right> \notag \\
			& + 4 y A_{ \chi_{\text{eff}} , \delta\chi}\left< \chi_{\text{eff}} \right>^2 + 2( \mu_2 \kappa_1 + \mu_1 \kappa_2 - 1 )\left< \chi_{\text{eff}} \right> \label{eq:eq:S00A-SS001-EQ0017}\\
			Q_{16} &= 2 G_{\delta\chi} ( 4y( A_{ \chi_{\text{eff}} , \chi_{\text{eff}} }\left< \chi_{\text{eff}} \right> + A_{ \chi_{\text{eff}} , \delta\chi}\left<\delta\chi \right> ) - \mu_1 \kappa_2 + \mu_2 \kappa_1 + \delta\mu ) \notag \\
			&+ 2 G_{\chi_{\text{eff}}} ( 4y( A_{ \chi_{\text{eff}} , \chi_{\text{eff}} }\left< \chi_{\text{eff}} \right> + A_{ \chi_{\text{eff}} , \delta\chi}\left<\delta\chi \right> ) + \mu_1 \kappa_2 + \mu_2 \kappa_1 - 1 ) \label{eq:eq:S00A-SS001-EQ0018}
			\\
			Q_{17} &= 4y ( A_{ \chi_{\text{eff}} , \delta\chi} ( G_{\delta\chi}^2 + G_{\chi_{\text{eff}}}^2  ) + 2 A_{ \chi_{\text{eff}} , \chi_{\text{eff}} } G_{\delta\chi} G_{\chi_{\text{eff}}} ) \label{eq:eq:S00A-SS001-EQ0019}
			.
		\end{align}
	\end{subequations}
	
	Using these definitions the entire derivative can be rewritten as	
	\begin{equation}\label{eq:eq:S00A-SS001-EQ020}
		\begin{split}
			\frac{d\phi_z}{dt} \approx& \frac{J y^6}{2} \left\{ Q_1+Q_2\sin(\psi) + \frac{ H_0 + H_1 \sin(\psi) + H_2 \sin^2(\psi) + H_3 \sin^3(\psi) }{ ( 1 + H_- \sin(\psi))( 1 + H_+ \sin(\psi))} \right\},
		\end{split}
	\end{equation}
	where
	\begin{subequations}
		\begin{align}
			\label{eq:eq:S00A-SS001-EQ021}
			H_0 &= \frac{ Q_3 Q_5 + Q_{11} + Q_{13} Q_{15} }{Q_8 Q_9}
			\\
			\label{eq:eq:S00A-SS001-EQ021}
			H_1 &= \frac{ Q_3 Q_6 + Q_5 Q_4 + Q_{12} + Q_{13}Q_{16} + Q_{14} Q_{15} }{Q_8 Q_9}
			\\
			\label{eq:eq:S00A-SS001-EQ021}
			H_2 &= \frac{ Q_3 Q_7 + Q_4 Q_6 + Q_{13} Q_{17} + Q_{14} Q_{16} }{Q_8 Q_9}
			\\
			\label{eq:eq:S00A-SS001-EQ021}
			H_3 &= \frac{ Q_4 Q_7 + Q_{14} Q_{17} }{Q_8 Q_9}
			\\
			\label{eq:eq:S00A-SS001-EQ021}
			H_- &= -\frac{Q_{10}}{Q_8}
			\\
			\label{eq:eq:S00A-SS001-EQ021}
			H_+ &= \frac{Q_{10}}{Q_9}.
		\end{align}
	\end{subequations}
	The angles appearing in the solutions for $\phi_z$ and $\zeta$ are related to the above expressions in the following way
	\begin{subequations}
		\begin{align}
			\label{eq:eq:S00A-SS001-EQ022}
			\Phi_0 &= \frac{J y^6}{2} \left\{ Q_1 + \frac{H_2 H_+ H_- - H_3 H_- - H_3 H_+ }{{H_+}^2 {H_-}^2}  \right\}
			\\
			\label{eq:eq:S00A-SS001-EQ023}
			\Phi_s &= \frac{J y^6}{2}\left\{ Q_2 + \frac{H_3}{H_+ H_-} \right\}
			\\
			\label{eq:eq:S00A-SS001-EQ024}
			\Phi_+ &= \frac{J y^6}{2} \left\{ \frac{H_0 {H_+}^3 - H_1 {H_+}^2 + H_2 {H_+} - H_3}{( H_+ - H_-) {H_+}^2 \sqrt{1-{H_+}^2 } } \right\}
			\\
			\label{eq:eq:S00A-SS001-EQ025}
			\Phi_- &= - \frac{J y^6}{2} \left\{ \frac{H_0 {H_-}^3 - H_1 {H_-}^2 + H_2 {H_-} - H_3}{( H_+ - H_-) {H_-}^2 \sqrt{1-{H_-}^2 }} \right\}
			\\
			\label{eq:eq:S00A-SS001-EQ026}
			\Theta_0 &= \frac{ 2L + \delta\mu \left<\delta\chi\right> + \left<\chi_{\text{eff}}\right> }{2J}
			\\
			\label{eq:eq:S00A-SS001-EQ027}
			\Theta_s &= \frac{ \delta\mu G_{\delta\chi} + G_{\chi_{\text{eff}}} }{ 2J} .
		\end{align}
	\end{subequations}

	\newpage
	\addcontentsline{toc}{chapter}{References}
	\bibliographystyle{unsrt}
	\bibliography{master}{}

\end{document}